\documentclass[12pt]{article}
\usepackage{epsfig,a4}
\usepackage{cite}
 \advance\oddsidemargin by -1in
 \advance\evensidemargin
 by -1in
 \marginparsep
 0.4cm
 \advance\topmargin by
 -0.0in
 \makeindex
 \def\Pom{{ I\!\!P}}
 \def\Reg{{ I\!\!R}}
 \def\gsim{\mathrel{\rlap{\lower4pt\hbox{\hskip1pt$\sim$}}
 \raise1pt\hbox{$>$}}}

 \newcommand\la{\langle}
 \newcommand\ra{\rangle}
 \newcommand\beq{\begin{equation}}
 
 \newcommand\eeq{\end{equation}}
 \newcommand\beqn{\begin{eqnarray}}
 \newcommand\eeqn{\end{eqnarray}}
\def\mb{\,\mbox{mb}}
\def\fm{\,\mbox{fm}}
\def\GeV{\,\mbox{GeV}}

\def\lsim{\mathrel{\rlap{\lower4pt\hbox{\hskip1pt$\sim$}}
    \raise1pt\hbox{$<$}}}         
\def\gsim{\mathrel{\rlap{\lower4pt\hbox{\hskip1pt$\sim$}}
    \raise1pt\hbox{$>$}}}         
\def\Re{\,\mbox{Re}\,}

\def\mb{\,\mbox{mb}}
\def\fm{\,\mbox{fm}}  
\def\GeV{\,\mbox{GeV}}
\def\MeV{\,\mbox{MeV}}
\def\sq{\sigma_{\bar qq}^N}
\def\st{\sigma_{tot}^{hN}}
\def\sel{\sigma_{el}^{hN}}
\def\sinhad{\sigma_{in}^{hN}}
\def\sdd{\sigma_{sd}^{hN}}

\def\sta{\sigma_{tot}^{hA}}
\def\sela{\sigma_{el}^{hA}}
\def\sina{\sigma_{in}^{hA}}
\def\stn{\sigma_{tot}^{NN}}
\def\seln{\sigma_{el}^{NN}}
\def\sinn{\sigma_{in}^{NN}}
\def\std{\sigma_{tot}^{pd}}

\def\sind{\sigma_{in}^{pd}}
\def\ssdn{\sigma_{sd}^{NN}}
\def\sddn{\sigma_{dd}^{NN}}

\def\sddd{\sigma_{dd}^{hN}}

\def\doublespace{\def\baselinestretch{1.6}\large\normalsize}
\def\normalspace{\def\baselinestretch{1.0}\normalsize}
\def\Caption#1{
  \normalspace
  \begin{quotation}\caption{\sl #1}\end{quotation}
  \doublespace
}
\begin{document}
\date{}

\title{\bf Transparent Nuclei and\\ Deuteron-Gold Collisions at
RHIC\footnote{ Based on lectures given by the author at Workshop on High
$p_T$ Correlations at RHIC, Columbia University, May-June, 2003.}}
 
\maketitle
 
\begin{center}
 
\vspace*{-1.5cm}
 {\large B.Z.~Kopeliovich}
 \\[0.5cm]
{\sl Max-Planck Institut f\"ur Kernphysik, Postfach 103980, 69029
Heidelberg}\\[0.2cm]
{\sl Institut f\"ur Theoretische Physik der Universit\"at, 93040
Regensburg} \\[0.2cm]
{\sl Joint Institute for Nuclear Research, Dubna, 141980 Moscow
Region, Russia}
\end{center}                                                              
 
\vspace{1cm}

\begin{abstract} 
 The current normalization of the cross section of inclusive high-$p_T$
particle production in deuteron-gold collisions measured at RHIC relies
on Glauber model calculations for the inelastic $dAu$ cross section.  
These calculations should be corrected for diffraction. Moreover, they
miss the Gribov's inelastic shadowing which makes nuclei more transparent
(color transparency) and reduce the inelastic cross section. The
magnitude of this effect rises with energy and one may anticipate it to
affect dramatically the normalization of the RHIC data. We evaluate the
inelastic shadowing corrections employing the light-cone dipole formalism
which effectively sums up multiple interactions in all orders. We found a
rather modest correction factor for the current normalization of data for
high-$p_T$ hadron production in $d-Au$ collisions. The results of
experiments insensitive to diffraction (PHENIX, PHOBOS) should be
renormalized by about $20\%$ down, while those which include diffraction
(STAR), by only $10\%$.  In spite of smallness of the correction it
eliminates the Cronin enhancement in the PHENIX data for pions. The
largest theoretical uncertainty comes from the part of inelastic
shadowing which is related to diffractive gluon radiation, or gluon
shadowing. Our estimate is adjusted to data for the triple-Pomeron
coupling and is small, however, other models do not have such a
restrictions and predict much stronger gluon shadowing.  Thus, one
arrives at quite diverse predictions for the correction factor which may
be even as small as $K=0.65$. Therefore, one should admit that the
current data for high-$p_T$ hadron production in $dAu$ collisions at RHIC
cannot exclude in a model independent way a possibility of initial state
suppression proposed by Kharzeev-Levin-McLerran. To settle this
uncertainty one should directly measure of the inelastic $d-Au$ cross
sections at RHIC. Also collisions with a tagged spectator nucleon may
serve as a sensitive probe for nuclear transparency and inelastic
shadowing.  We found an illuminating quantum-mechanical effect: the
nucleus acts like a lens focusing spectators into a very narrow cone.

 \end{abstract}


\newpage


\tableofcontents
 
\vspace*{1cm}

\section{Introduction}

Recent data for high-$p_T$ hadron production in deuteron-gold collisions
at $\sqrt{s}=200\GeV$ at RHIC \cite{phenix,star,phobos} demonstrate
importance of these measurements for proper interpretation of data from
heavy ion collisions. The observed nuclear effects at high-$p_T$ are
pretty weak, the enhancement (Cronin effect) measured for pions by PHENIX
is only about $10-20\%$, in accordance with expectation of \cite{knst}
and with somewhat larger effect found in \cite{miklos}, while a
suppression, rather than enhancement was predicted in \cite{klm}. To
discriminate between these predictions the data should have at least few
percent accuracy.

In this notes we draw attention to the fact that only the shape of
$p_T$-distribution was measured experimentally, while the normalization
of the data is based on theoretical calculations which are not correct.
Therefore, the reported results of deuteron-gold measurements
\cite{phenix,star,phobos} 
may be altered by more appropriate calculations.

The nucleus to nucleon ratio demonstrating the well known Cronin effect 
\cite{cronin} is defined as,
 \beq
R_{A/N}(p_T) = \frac{d\sigma^{hA}/d^2p_T}
{A\,d\sigma^{hN}/d^2p_T}\ .
\label{2}
 \eeq
 At large $p_T$, of the order of few $\GeV$ this ratio exceeds one, 
but eventually approaches one at very high $p_T$ as is expected according 
to $k_T$-factorization (it may even drop below one due to the EMC effect 
at large Bjorken $x$).

Absolute values of the high-$p_T$ nuclear cross sections are difficult to
measure at RHIC, only the fraction of the total inelastic cross section,
$dN^{hA}/d^2p_T$ is known. Then, one has to normalize it multiplying the
fraction by the total inelastic cross section,
 \beq
R_{A/N}(p_T) = 
\frac{\sina\,dN^{hA}/d^2p_T}
{A\,\sinn\,dN^{hN}/d^2p_T}=
\frac{1}{N_{coll}}\ 
\frac{dN^{hA}/d^2p_T}
{dN^{hN}/d^2p_T}\ ,
\label{4}
 \eeq
 where
 \beq
N_{coll} = A\,\frac{\sigma^{hN}_{in}}{\sigma^{hA}_{in}}\ ,
\label{10}
 \eeq
 In some experiments the denominator in (\ref{2}), $d\sigma^{hN}/d^2p_T$
was directly measured or borrowed from other measurements, otherwise it
should be corrected for diffractive dissociation of the colliding protons
which possesses a large rapidity gap and escapes detection. In what
follows we assume that the denominator in (\ref{2}),
$d\sigma^{hN}/d^2p_T$ was directly measured (see, however, discussion in
Sect.~\ref{elasticity}) and concentrate on nuclear effects, i.e. the
inelastic nuclear cross section $\sigma^{NA}_{in}$ which was calculated
in \cite{phenix,star,phobos} in an oversimplified approach.

\subsection{Number of collisions: who is actually colliding?}

The Glauber approach is a model for the elastic hadron-nucleus amplitude.
It is demonstrated in Appendix~\ref{appendA} how to calculate inelastic
and quasielastic cross section using unitarity and completeness.  The
model does not say anything about exclusive channels of inelastic
interaction. One can formally expand the Glauber exponential, and it
looks like a series corresponding to different numbers of inelastic
collisions of the same hadron and with the same inelastic cross section.
However, a high-energy hadron cannot interact inelastically many times,
since the very first inelastic collision breaks down coherence between
the constituents of the hadron. It takes a long time proportional to the
energy to produce a leading hadron in final state.

The cross section of inelastic hadron-nucleus collision, $\sina$, is
related to the probability for the incoming hadron to get the very first
inelastic collision, usually on the nuclear surface. This is why
$\sina\propto A^{2/3}$. Since the process is fully inclusive, subsequent
final state interactions do not affect the cross section due to
completeness.  

$N_{coll}$ defined via expansion of the Glauber exponential term should
not be treated as multiple sequential interactions of the projectile
hadron (like expansion of the exponential describing the time dependence
of particle decay does not mean that the particle can decay many times).
After the first inelastic interaction the debris of the projectile hadron
keep traveling through the nucleus, but their interactions apparently
have little to do with the properties of the incoming hadron and its
inelastic cross section. Formally, one can relate $N_{coll}$ to the mean
number of the Pomerons which undergo unitarity cuts. The
Abramovsky-Gribov-Kancheli (AGK) cutting rules \cite{agk} which are not
proven in QCD, assume that these cuts have the same eikonal weights as
given by the Glauber model. In this approach multiple interactions are
not sequential (planar), but occur in parallel, i.e. they allow a
simultaneous unitarity cut. In terms of the light-cone approach multiple
interactions correspond to higher Fock states in the projectile hadron.
The constituents of this states propagate through the nucleus and
experience their first inelastic interactions independently of each
other. The probability of such multiple interactions has little to do
with the properties of the low Fock states which dominate the
hadron-nucleon cross section, therefore it should not be expressed as a
power of $\sinhad$.  At any rate, whether the AGK weights are correct or
not, it is clear $N_{coll}$ cannot be treated as sequential interactions
of the projectile hadron.

The first inelastic collision of the incoming hadron is a soft
color-exchange interaction. The projectile partons do not alter either
their number (for a given Fock state), or their longitudinal momenta, but
the whole system of partons acquire a color. Therefore, the remnants of
the hadron turn out to be color connected to the remnants of the target.
Then new partons are produced from vacuum (e.g. via the Schwinger
mechanism) aiming to neutralize the color of the projectile partons.
Their momenta are much smaller than of the projectile partons. Such an
excited and color neutral partonic system keeps propagating through the
medium and experiencing new soft color exchange interactions similar to
ordinary hadrons.  The corresponding cross section is subject to color
screening and is controlled by the transverse, rather than longitudinal
size of the system.

From the practical point of view, there is nothing wrong in using
$N_{coll}$ as a multiplication factor for hard reactions, since within
the Glauber model it is proportional to the nuclear thickness function
$T^N_A(b)$, i.e. to the number of opportunities for a parton to perform a
hard process.  Indeed, the projectile high energy partons participate in
hard reactions independently of the accompanying partons, since color
screening plays no role for a hard interaction.  Moreover, naively, one
may expect that this factor $T^N_A(b)$ ($T_{AB}$ in the case of $AB$
collision) is all one needs to normalize a hard process, and this
normalization is independent of the soft cross section
$\sigma_{in}^{hN}$. However, $N_{coll}$ is defined for events where
inelastic collision did happen. Therefore, it must be properly normalized
by the probability for the incoming hadron to make inelastic interaction
at the given impact parameter,
 \beq
N_{coll}(b) = \frac{\sinn\,T^N_A(b)}
{1-\exp[-\sinn\,T^N_A(b)]}\ .
\label{50}
 \eeq
 Averaging this expression over inelastic collisions at different
impact parameters one indeed arrives to the expression Eq.~(\ref{10}).

\subsection{Correcting data for \boldmath$R_{d-Au}$}

The current analyses of RHIC data \cite{phenix,star,phobos} calculate
$N_{coll}$ in Glauber Monte Carlo model assuming $\sinn=41-42\mb$. In
these notes we challenge these calculations and show that the published
results for $d-Au$ collisions are subject to important corrections and
the conclusions are model dependent.

There are two major corrections to be done to the Cronin ratio
Eq.~(\ref{4}) measured at RHIC.  We combine them in a correction factor
$K$,
 \beq 
R_{dA}(p_T) = R^{RHIC}_{dA}(p_T)
\times K\ , 
\label{6}
 \eeq
 where
 \beq
K = K_{Gr}\times K_{Gl}\ .
\label{8}
 \eeq
 Here $K_{Gr}$ is the correction related to Gribov's inelastic
shadowing missed in Glauber model calculations. It is introduced in
Sect.~\ref{in-sh} and calculated throughout the paper.

Even within the Glauber model the calculations performed in
\cite{phenix,star,phobos} should be corrected by a factor $K_{Gl}$. It
originates from a more accurate treatment of the inelastic $NN$ cross
section which should correspond to the class of events selected for the
analysis, as is explained in Sect.~\ref{elasticity}. This correction is
calculated in Sect.~\ref{deuteron-nucleus}.

There is an additional correction which should be included into (\ref{8})
if one needs to compare with theoretical predictions for the Cronin
effect for $pA$ collisions. It is related to the fact that the deuteron
is a nucleus and is also subject to the Cronin effect. Therefore
high-$p_T$ enhancement in $d-A$ must be somewhat stronger than in $p-A$
collisions. This correction is evaluated in Sect.~\ref{cronin} and
found rather small.

\subsubsection{Inelastic shadowing.}\label{in-sh}

\begin{figure}[tbh]
\includegraphics{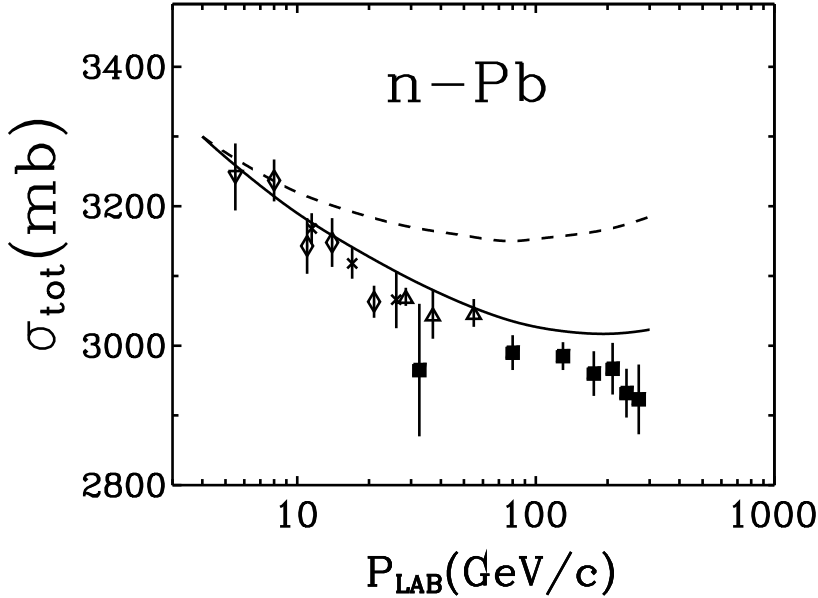}
\begin{center}
\vspace{8.5cm}
\parbox{13cm}
{\caption[Delta]
 {\sl Data and calculations \cite{murthy} for the total neutron-lead
cross section as function of energy. The dashed curve corresponds to the
Glauber model, while the solid curve is corrected for Gribov's inelastic
shadowing.}
 \label{murthy}} 
\end{center}
 \end{figure}

It is known that Gribov's inelastic corrections \cite{gribov} to the
Glauber approximation make nuclear matter more transparent and reduce the
hadron-nucleus cross sections compared to the Glauber model. This effect
steeply rises with energy, as one can see from the example depicted in
Fig.~\ref{murthy} for the total neutron-lead cross section measured and
calculated in \cite{murthy}.
 Apparently, the Glauber model overestimates the cross section, and the
deviation rises with energy.  Without a good theoretical input one cannot
predict what will happen at the energy of RHIC which is 100 times higher
than in fixed-target experiments at Fermilab. This is a serious challenge
for the theory to calculate the inelastic $dA$ cross section at these
energies, and the results apparently will be model dependent. However, it
is certain that the sign of the correction remains negative and it can
only rise with energy, i.e. cannot be smaller than what is shown in
Fig.~\ref{murthy} for low energies.

Our own estimates summarized in Table~\ref{tab} give a moderate
reduction, about $20\%$. The weakness of the effect is based on a proper
treatments of diffraction and is fixed by data on large mass diffractive
dissociation of protons \cite{kst2}.  At the same time, many models
predict quite a strong gluon shadowing even at high virtualities.
Naturally, this effect should not be weaker in soft $NN$ interactions.
Then it may lead to a stronger suppression of the inelastic $dA$ cross
section than we found, as it is discussed in Sect.~\ref{models}.

Note that although inelastic shadowing makes nuclear medium more
transparent, the mean number of collisions increases according to
(\ref{10}). It sounds counter-intuitive that a hadron experiences more
collisions in a less absorptive medium. Formally it follows from
{\ref{10}), but can be explained qualitatively. For instance, if one
calculated the mean number of collisions in a photoabsorption reaction on
a nucleus using the Glauber formula, the result will be very small,
proportional to $\alpha_{em}$. However, $N_{coll}$ is defined for events
when inelastic collision took place. In this case it comes from hadronic
fluctuations of the photon and is much larger than number of collisions
given by the Glauber formulas Eq.~(\ref{10}). This example
explains why $N_{coll}$ increases due to inelastic shadowing.

\subsubsection{How inelastic is the inelastic cross section?} 
\label{elasticity}

As far as one needs to calculate the deuteron-nucleus inelastic cross
section it should be done in correspondence with the class of events
selected by the trigger. The cross section calculated via Glauber
Monte-Carlo generator in all three experiments corresponds to the Glauber
formula derived in Appendix~A, Eq.~(\ref{a.100}), where $\sinn$ is the
total inelastic $NN$ cross section. Then, according to derivation,
Eq.~(\ref{a.100}) describes the total inelastic cross section on a
nucleus minus the part related to quasielastic nuclear excitations (with
no hadron produced). This is not what was actually measured in any of the
three experiments \cite{phenix, star, phobos}. These experiments have
different event selections and the calculations should comply with that.

The STAR experiment triggers on forward neutrons from the gold
\cite{star} and detects all inelastic $d-Au$ collisions including
quasielastic excitation of the gold\footnote{I appreciate the very
informative communication with Carl Gagliardi on this issue.}. In this
case, according to the Glauber formalism presented in Appendix~A, one
should rely on Eq.~(\ref{a.65}) with $\stn=51\mb$, rather than
$\sinn=42\mb$. At the same time, the two other spectrometers, PHENIX and
PHOBOS, seem to be insensitive to large rapidity gap events, i.e.
diffractive excitations of the deuteron and gold
\cite{phenix,phobos}\footnote{A part of diffraction might have been
included into the trigger efficiency of the PHENIX spectrometer. Namely,
double diffraction, i.e. excitation of nucleons in both the deuteron and
gold can reach and fire sometimes the closest of the two BBC triggers
covering pseudorapidity intervals $\eta = \pm(3-3.9)$. However, the main
part of diffraction, single diffractive excitation of either the
deuteron, or gold, hardly could reach the opposite hemisphere and fire
both the BBC triggers simultaneously, what is the trigger condition. I am
thankful to Barbara Jacak and Sasha Milov for informative and clarifying
discussions of this issue.}. Then Eq.~(\ref{a.100}) should be applied
with a replacement $\sinn\Rightarrow \sinn-2\ssdn-\sddn \approx 30\mb$,
i.e. the single and double diffraction must be subtracted (see details in
Sect.~\ref{deuteron-nucleus}) \cite{dino,cdf}. Apparently, it makes
difference whether one performs calculations with input cross section
$51\mb$, or $42\mb$, or $30\mb$.

The numerator in (\ref{50}), $\sinn$ is even more sensitive than the
denominator to assumptions which inelastic channels should be included.
However this does not seem to be a problem, since the cross section of
high-$p_T$ production in $pp$ collisions, $d\sigma^{pp}/d^2p_T$, was
directly measured in the all three experiments \footnote{Although it is
stated in \cite{phenix} that the cross section $d\sigma^{pp}/d^2p_T$ was
normalized to $42\mb$, in fact it was measured \cite{barbara,mike}.}, and
we consider only the nuclear modification factor Eq.~(\ref{6}) in what
follows.

\subsection{The outline}\label{outline}

This paper is organized as follows. We present a brief and simple
derivation of basic formulae of the Glauber model \cite{glauber} in
Appendix~\ref{appendA}.  In Sect.~\ref{deuteron-nucleus} we treat the
deuteron as a nucleus and generalize the Glauber model for this case. We
derive formulae for the cross sections of different channels, perform
numerical calculations and present the results in Table~\ref{tab}. We
corrected the input inelastic $NN$ cross section for diffraction and
found a smaller $\sigma_{in}^{dA}$ than in \cite{phenix,phobos}, but
larger than in \cite{star}.

Events with a tagged spectator nucleon may serve as a sensitive probe for
nuclear transparency, since the spectator must propagate through the
nucleus with no interaction. We calculate the total cross section for
this channel and the transverse momentum distribution of the spectators.
On the contrary to naive expectation that noninteracting nucleons retain
their primordial Fermi momentum distribution, we found an amazingly
strong focusing effect. Namely, the nucleus acts like a lens focusing the
spectators into a narrow cone with momentum transfer range of the order
of the inverse nuclear radius. The transverse momentum spectrum of the
spectators acquires typical diffraction structure having minima and
maxima.

Inelastic shadowing corrections are introduced in Sect.~\ref{in-corr}.
First, we use the traditional presentation in terms of inelastic 
diffractive excitations in intermediate state of hadron-nucleus elastic 
amplitude (Sect.~\ref{sect-kk}). This approach is quite restricted, being 
unable to deal with higher order scattering terms which are especially 
important at high energies. Therefore, we switch to the eigenstate 
representation introduced in general terms in Sect.~\ref{eigen}.
Its realization in QCD is the light-cone color-dipole approach presented 
in Sect.~\ref{dipole}. 

The part of the inelastic corrections related to the lowest hadronic Fock
component consisted only of valence quarks corresponds to diffractive
excitation of resonances in usual terms. This contribution is analyzed
and estimated numerically in Sect.~\ref{quarks}. We demonstrate that this
corrections make heavy nuclei much more transparent: instead of
exponential attenuation we found a linear dependence on the inverse
nuclear thickness (Sect.~\ref{transparency}). Correspondingly, we derived
formulae for cross sections of different channels corrected for inelastic
shadowing for hadron-nucleus (Sect.~\ref{Xsection}), and deuteron-nucleus
(Sect.~{dA}) collisions. In Sect.~\ref{real} we study possibility to
improve our calculations. We tested sensitivity of our results to the
form of the nucleon wave function, and derived formulae for the case of
a realistic saturated dipole-nucleon cross section.

Gluonic excitations corresponding to Fock states containing extra gluons
are considered in Sect.~\ref{gluons}. They correspond to diffractive 
excitations of large mass which are known to have quite a small cross 
section. This smallness leads to a prediction of rather weak
gluonic shadowing, $\sim 20\%$, and small contribution to the 
inelastic corrections. At the same time other models predict much
stronger gluon shadowing (Sect.~\ref{models}) which may substantially
change the normalization of the $dAu$ data.

Since nuclear matter become more transparent due to inelastic shadowing,
the number of participants changes as well. In Sect.~\ref{npart} we found
this effect to be sizeable.

In Sect.~\ref{cronin} we sum up the effects considered so far to see how
much they affect the $dAu$ data. The results are presented in
Table~\ref{tab}. We also corrected the PHENIX data for high-$p_T$
pions to see how important are these corrections compared to the
current error bars. We found a considerable change: the Cronin effect
for high-$p_T$ pions disappeared.

Our observations are summarized in Sect.~\ref{summary}. The main 
conclusion is that the current data for high-$p_T$ hadron production in 
deuteron-gold collisions are not decisive, and should be complemented 
with direct measurements of the inelastic $dAu$ cross section.

\section{Extending the Glauber model to deuteron-nucleus
collisions}\label{deuteron-nucleus}

The basic formulae of Glauber model for hadron-nucleus collisions are presented
in Appendix~\ref{appendA}. If to treat the deuteron as a hadron, one can
calculate the $dA$ total, total elastic and inelastic cross section provided
that the elastic $dN$ amplitude is known. The latter can be calculated
employing the Glauber model too. This is done in Appendix~\ref{appendA}.

One can do calculations differently, treating the deuteron as a system of 
two nucleons interacting with the nucleus. In this case one can consider
more reaction channels as deuteron excitation etc., which are missed in
the previous approach.

\subsection{The total cross section}.
We generalize Eq.~(\ref{a.50}) from Appendix~\ref{appendA} for a deuteron
beam as 
following,
 \beqn
\sigma_{tot}^{dA} &=& 
2{\rm Re}\int d^2r_T\,\left|\Psi_d(r_T)\right|^2
\nonumber\\ &\times&
\left\la 0\left|1 - \prod_{k=1}^A\left[1-
\Gamma^{pN}(\vec b-\vec r_T/2-\vec s_k)\right]\left[1-
\Gamma^{nN}(\vec b+\vec r_T/2-\vec s_k)\right]
\right|0\right\ra
\nonumber\\ &=& 2\int d^2b\,
\int d^2r_T\,\left|\Psi_d(r_T)\right|^2
\left\{1 - \exp
\left[-{1\over2}\,\stn\,\left(T_A^N(\vec b+{1\over2}\vec r_T)
+ T_A^N(\vec b-{1\over2}\vec r_T)\right)
\right.\right.\nonumber\\ &+&
\left.\left. \seln\,T_A^{N}(b)\,
\exp\left(-\frac{r_T^2}{4B_{NN}}\right)
\right]\right\}\,
\label{38}
 \eeqn
 where $\vec r_T$ is the transverse nucleon separation in the deuteron;
$|\Psi_d(r_T)|^2$ is the deuteron light-cone wave function squared and
integrated over relative sharing by the nucleons of the deuteron
longitudinal momentum. It is presented in Appendix~\ref{appendC}. The
effective
nuclear thickness function, $T^N_A(b)$, convoluted with the $NN$ elastic
amplitude is introduced in (\ref{a.51}).

We did calculations with nuclear density in the Woods-Saxon form
 \beq
\rho_A(r)= \frac{3\,A}{4\pi R_A^3(1+\pi^2a^2/R_A^2)}\ 
\frac{1}{1+\exp\left(\frac{r-R_A}{a}\right)}
\label{36}
 \eeq
 with $R_A=6.38\fm$ and $a=0.54\fm$, same as in \cite{phenix} for easier
comparison.  The result for the total cross section $\sigma^{dAu}_{tot}$
is shown in Table~\ref{tab}.
 \begin{table}
\Caption{
 \label{tab}
 Results for different cross sections and numbers of collisions
calculated using Glauber approximation (Sect.~\ref{deuteron-nucleus}),
corrected for inelastic shadowing related to valence quark fluctuations
(Sect.~\ref{quarks}), and for gluon shadowing (Sect.~\ref{gluons}). The
results including the ultimate renormalization factor $K$ depend on the
experimental set up and are different for the STAR and PHENIX
experiments}
 \vskip3mm  
\begin{center} 
\begin{tabular}{|c|c|c|c|c|c|} 
\hline
\vphantom{\bigg\vert}
  & Observable
  & Glauber
  & Valence quark                 
  & Plus gluonic 
  & Correction
\\[-0.4cm]
&&model&fluctuations & excitations & factor 
\\
\hline &&&&&\\[-8mm]
&$\sigma^{dAu}_{tot}[\mb]$ & 4110.1  & 3701.0  & 3466.2&
\\
\hline &&&&&\\[-6mm]
S&$\sigma^{dAu}_{in}[\mb]$ & 2422.7  & 2226.6(2335.8)  & 2118.3(2228.3)&
\\
T&Factor $K$ in (\ref{6})-(\ref{8})& $K_{GL}=$1.04&  &
$K_{Gr}=0.87(0.92)$&K=0.91(0.96)
\\
A&$N_{coll}^{in}(min.b.)$ & $6.9$ & $7.5$ & $7.9$&\\
R&$\sigma^{dAu}_{in}(tagg)[\mb]$ & 458.4 & 544.9(511.5) & 551.8(520.1)& 
\\ 
&$N_{coll}^{in}(tagg)$ & $2.9$ & $4.4$ & $5.0$& \\
[3mm]\hline &&&&&\\[-6mm]
P&$\sigma^{dAu}_{non-diff}[\mb]$ & 2146.0 & 1998.3(2100.1) &
1930.3(2033.7)&
\\
H&Factor $K$& $K_{Gl}=0.92$&  & $K_{Gr}=0.9(0.95)$&K=0.83(0.87)
\\
E&$N_{coll}^{non-diff}(min.b.)$ & $5.5$ & $5.9$ & $6.1$&\\
N&$\sigma^{dAu}_{non-diff}(tagg)[\mb]$ & 324.3 & 480.2(451.5) &
498.4(470.6)&
\\
I&$N_{coll}^{non-diff}(tagg)$ & $2.3$ & $2.9$ & $3.2$&
\\
X&&&&&\\
[1mm]
\hline   
\end{tabular}
\end{center}
\end{table}

Eq.~(\ref{38}) is easy to interpret. The two first term in the exponent
correspond to independent interaction of two nucleons separated by
transverse distance $\vec r_T$. Of course, the smaller $r_T$ is, the
stronger nucleons shadow each other, and this is accounted for by the
third term.

 One can see the difference between this expression and Eq.~(\ref{a.50})
(for $h\equiv d$). In the latter cases the averaging over $\vec r_T$ is
put up into the exponent, while in the former case, Eq.~(\ref{38}), the
whole exponential is averaged. We will see in Sect.~\ref{eigen} that this
difference is a part of the Gribov's inelastic corrections, so (\ref{38})
makes the first step beyond the Glauber approximation.  

Note that the last term in the exponent in (\ref{38}) is quite small.
Besides smallness of $\seln/\stn$, the exponential factor is
rather small.  The mean value of the exponent is $\la r_T^2\ra/4B_{NN}
\approx 5$. This term reduces the total $d-Au$ cross section by $1.3\%$
only.

\subsection{The cross section of elastic \boldmath$dA$ scattering and
deuteron breakup \boldmath$dA\to pnA$}.

According to (\ref{a.60}) in order to find elastic $dA$ cross section one
should square the partial elastic amplitude and integrate over $b$,
 \beqn 
\sigma^{dA}_{el}&=&
\int d^2b\, \left|
\int d^2r_T\,\left|\Psi_d(r_T)\right|^2 
\left\{1 - \exp \left[-{1\over2}\stn
\Biggl(T_A^N(\vec b+{1\over2}\vec r_T)
+ T_A^N(\vec b-{1\over2}\vec r_T)\Biggr) 
\right.\right.\right.\nonumber\\ &+&
\left.\left.\left. \seln\,T_A^{N}(b)\, 
\exp\left(-\frac{r_T^2}
{4B_{NN}}\right)
\right]\right\}\right|^2\, 
\label{39.0}
 \eeqn

This is the square of the average of the elastic amplitude
over deuteron configurations. If, however, to take average of the
amplitude squared, that will include also dissociation $d\to pn$, i.e.
 \beqn 
\sigma^{dA}_{el}(dA\to dA) + 
\sigma^{dA}_{diss}(dA\to npA) &=&
\int d^2b\, \int d^2r_T\,\left|\Psi_d(r_T)\right|^2 
\nonumber\\ &\times&
\left|1 - \exp \left[-{1\over2}\stn
\Biggl(T_A^N(\vec b+{1\over2}\vec r_T)
+ T_A^N(\vec b-{1\over2}\vec r_T)\Biggr) 
\right.\right.\nonumber\\ &+& \left.\left. 
\seln\,T_A^{N}(b)\, \exp\left(-\frac{r_T^2}
{4B_{NN}}\right)
\right]\right|^2\, 
\label{39}
 \eeqn

\subsection{The total inelastic cross section} 

Subtracting from the total cross section the elastic part one gets the
cross section of all inelastic channels in $d-A$. We, however, prefer to
subtract the deuteron quasielastic breakup too, since it is not detected
by any of the RHIC experiments. Then we have,
 \beqn
\sigma^{dA}_{in} &=&
\sigma^{dA}_{tot}-\sigma^{dA}_{el}-
\sigma^{dA}_{diss}(dA\to npA) = 
\int d^2b\,
\int d^2r_T\,\left|\Psi_d(r_T)\right|^2
\nonumber\\ &\times&
\left\{1 - \exp
\left[-\stn\,\left(T_A^N(\vec b+{1\over2}\vec r_T)
+ T_A^N(\vec b-{1\over2}\vec r_T)\right)
\right.\right.\nonumber\\ &+&\left.\left.
2\,\seln\,T_A^{N}(b)\,
\exp\left(-\frac{r_T^2}{4B_{NN}}\right)
\right]\right\}\ .
\label{39a}
 \eeqn
 The result of numerical calculation for this cross section is exposed in
Table~\ref{tab}. This cross section covers diffractive  excitations as
well, therefore we place the result in the upper panel of the table
which is supposed to be related to experiments sensitive to diffraction
(STAR).

 The impact parameter distribution of the inelastic cross section is
plotted by dashed curve in Fig.~\ref{results-in}, and the integrated
cross section is shown in Table~\ref{tab}.
 \begin{figure}[tbh]
\includegraphics{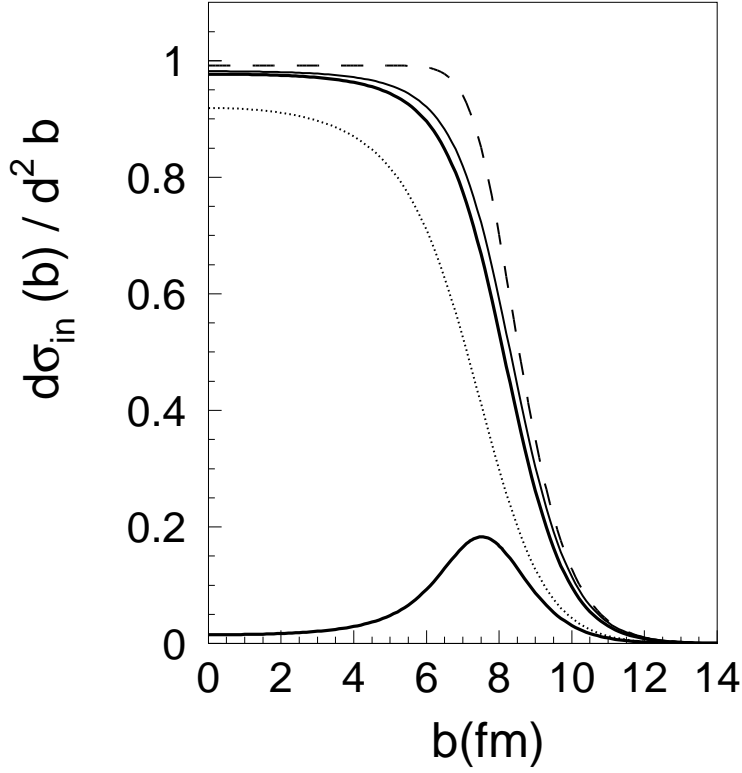}
 \begin{center}
 \vspace{10.5cm}
 \parbox{13cm} 
{\caption[Delta]
 {\sl The impact parameter distribution of inelastic deuteron-gold
collisions (three upper curves) including diffractive excitations (STAR
trigger). Impact parameter $\vec b$ corresponds to center of gravity of
the deuteron. The dashed curve corresponds to the Glauber approximation
Eq.~(\ref{39}). The thin solid curve include inelastic shadowing related
to excitation of the valence quark skeleton, Eq.~(\ref{63}). The thick
solid curve is final, it includes gluon shadowing as well.  The bottom
solid thick curve shows the difference between the Glauber and final
curves. The dotted curve shows the range of model uncertainty and
corresponds to gluon shadowing with $R_G=03$ (see Sect.~\ref{models}).
All curves are calculated with total cross section $\tilde\stn=51\mb$.}
 \label{results-in}} 
\end{center}
 \end{figure}

\subsection{The cross section of nondiffractive channels}.

In experiments insensitive to large rapidity gap event
one should employ the inelastic cross section
with all diffractive contribution is removed. Namely, one
should also subtract the cross sections of quasielastic excitation of the
nucleus, $A\to A^*$, and diffractive excitation of colliding nucleons.

The cross section of single ($dA\to dA^*$) and double ($dA\to pdA^*$)
quasielastic and quasidiffractive nuclear excitation reads [compare with
(\ref{a.90})],
 \beqn
&& \sigma_{qel}^{dA}(dA\to dA^*) +
\sigma_{qsd}^{dA}(dA\to pnA^*) =
\int d^2r_T\,\left|\Psi_d(r_T)\right|^2
\nonumber\\ &\times&
\left\{\left\la 0\left|\left|1 - \prod_{k=1}^A\left[1-
\Gamma^{pN}(\vec b-\vec r_T/2-\vec s_k)\right]\left[1-
\Gamma^{nN}(\vec b+\vec r_T/2-\vec s_k)\right]
\right|^2\right|0\right\ra
\right.\nonumber\\ &-&\left.
\left\la 0\left|1 - \prod_{k=1}^A\left[1-
\Gamma^{pN}(\vec b-\vec r_T/2-\vec s_k)\right]\left[1-
\Gamma^{nN}(\vec b+\vec r_T/2-\vec s_k)\right]
\right|0\right\ra^2\right\}
\nonumber\\ &=&                    
\int d^2b\,
\int d^2r_T\,\left|\Psi_d(r_T)\right|^2
\left\{\exp
\left[-\sinn\,\left(T_A^N(\vec b+{1\over2}\vec r_T)
+ T_A^N(\vec b-{1\over2}\vec r_T)\right)
\right.\right.\nonumber\\ &+& \left.\left.
4\,\seln\,T^N(b)\,\gamma(r_T)\,
\exp\left(-\frac{r_T^2}{4B_{NN}}\right)
\right] 
\right.\nonumber\\ &-& \left.
\exp\left[-\stn\left(T_A^N(\vec b+{1\over2}\vec r_T)
+ T_A^N(\vec b-{1\over2}\vec r_T)\right) +
\seln T^N(b)\exp\left(-\frac{r_T^2}{4B_{NN}}\right)
\right]\right\},
\label{39b}
 \eeqn
 where 
 \beq
\gamma(r) = 1-
{8\over3}\,\frac{\seln}{\stn}\,
\exp\left(-\frac{r_T^2}{8B_{NN}}\right) +
8\left(\frac{\seln}{\stn}\right)^2
\exp\left(-\frac{r_T^2}{4B_{NN}}\right)
\ ,
\label{39bb}
 \eeq
 is a correction factor hardly different from one. In what follows we do
not keep the small terms in (\ref{39bb}). Note that
the form of Eq.~(\ref{39b}) is analogous to that of Eq.~(\ref{a.90}).

For experiments insensitive to diffraction the quasielastic cross section
Eq.~(\ref{39b}) should be subtracted (\ref{39a}) and the result will be
similar to Eq.~(\ref{a.100}).
However, we still miss the contribution of diffractive channels
related to diffractive excitations of nucleons in the deuteron and 
nucleus. It is impossible to introduce consistently diffraction in the
framework of the Glauber model which is a single channel approximation.
Diffraction naturally emerges in the multiple coupled channel approach or
in the eigenstate method introduced below. Meanwhile, one can use the
following prescription.

Let us expand the exponentials in (\ref{39b}) in small expansion
parameter $\seln\,T_A(b)$ up to the first order,
 \beqn
\sigma_{qel}^{dA} &\approx&
\int d^2b\,
\int d^2r_T\,\left|\Psi_d(r_T)\right|^2\,
\exp\left[-\stn\,\left(T_A^N(\vec b+{1\over2}\vec r_T)
+ T_A^N(\vec b-{1\over2}\vec r_T)\right)\right]
\nonumber\\ &\times&
\seln\,\left[T_A^N(\vec b+{1\over2}\vec r_T) + 
T_A^N(\vec b-{1\over2}\vec 
r_T)\right] + ...
\label{39c}
 \eeqn

In order to include the possibility of diffractive excitation 
of nucleons in the colliding nuclei, one should replace in (\ref{39c})
 \beq
\seln \Rightarrow \tilde\seln = \seln + 2\ssdn + \sddn\ .
\label{39d}
 \eeq
 in all orders of $\seln\,T_A(b)$.
 This is a substantial correction since at RHIC energy $\seln=9\mb$,
and $\seln + 2\ssdn + \sddn = 21\mb$. 

The final Glauber model expression for the non-diffractive inelastic $dA$ 
cross section reads,
 \beqn
\sigma^{dA}_{non-diff} &=& \int d^2b\,
\int d^2r_T\,\left|\Psi_d(r_T)\right|^2
\left\{1 - \exp\left[-\tilde\sinn\,
\left(T_A^N(\vec b+{1\over2}\vec r_T)
+ T_A^N(\vec b-{1\over2}\vec r_T)\right)
\right.\right.\nonumber\\ &+&
\left.\left. 4\,\tilde\seln\,T_A^{NN}(b)\,
\exp\left(-\frac{r_T^2}{4B_{NN}}\right)
\right] \right\}\,
\label{39e}
 \eeqn
 where
 \beq
\tilde\sinn = \stn-\tilde\seln
\label{39f}
 \eeq

We calculated the non-diffractive part, Eq.~(\ref{39e}), of the inelastic
$d-Au$ cross section, and the result is shown in Table~\ref{tab}. The
corresponding number of collisions also presented in the Table is rather
small compared to the one quoted in \cite{phenix}. This is mainly due to
a smaller inelastic cross section $\tilde\sinn$ we use. 

\section{Quantum mechanics at work: illuminating focusing effect for
spectators}\label{tagged}

Assume that only the proton in the deuteron interacts inelastically with
the nucleus, while the neutron is a spectator (of course all following
results are symmetric relative to interchange $p \leftrightarrow n$).
This is a very interesting process of simultaneous interaction and no
interaction. It provides direct information about nuclear transparency.
Apparently, this process pushes the neutron to the ultra-periphery of the
nucleus where its survival probability is high, while the proton prefers
to hit the dense area of the nucleus and interact.

Naively, the survived spectator neutrons should maintain their primordial
transverse momentum distribution controlled by the deuteron size. This is
assumed in the Glauber Monte Carlo. However, quantum mechanics is at
work, and the nucleus acts like a lens focusing spectator neutrons. The
survival probability modifies the shape of the wave packet of the
spectators in the impact parameter plane. Correspondingly, their
$p_T$-distribution changes. This is how elastic scattering on an
absorptive target happens: it is not due to transparency of the target,
but is caused by absorption. In the limit of a completely transparent
target, the incoming plane wave is not disturbed and no scattering
occurs. Absorption makes a hole in the plane wave, and one can think
about the outside area of the incoming wave which undergoes elastic
scattering. On the other hand, one can subtract the incoming plane wave
whose Fourier transform is just a delta function (zero angle scattering)
and the rest is a wave packet with a transverse area of the target size.
A Fourier transform of this wave packet gives the elastic amplitude
[compare with (\ref{a.60})].

Thus, the spectator neutrons experience elastic scattering on the target,
rather than simply propagate with the undisturbed primordial transverse
Fermi momentum. Below we derive formulas which show how elastic
scattering of the spectator neutrons happens, and perform numerical
evaluation of the effect.

We start with the cross section of this process which can be written as,
 \beqn
\sigma_{tagg}^{dA}(dA\to nX) &=& 
{\rm Re}\int d^2r_T\,\left|\Psi_d(r_T)\right|^2
\left\la 0\left|\prod_{k=1}^A\left[1-
\Gamma^{nN}\left(\vec b-\vec r_T-\vec 
s_k\right)\right]^2\right.\right.
\nonumber\\ &\times& \left.\left.
\left\{1-\prod_{k=1}^A\left[1-
2\,\Gamma^{pN}\left(\vec b-\vec
s_k\right)\right]\right\}
\right|0\right\ra\ ,
\label{40}
 \eeqn 
 where $\vec b$ is the impact parameter of the proton.
The first factor here would be the elastic neutron-nucleus cross
section, if it were not weighted by the second term which is the
inelastic proton-nucleus cross section, i.e. the difference between the
total and elastic and quasielastic cross sections (see
Appendix~\ref{appendA}).

After integration over the coordinates of bound nucleons we get,
 \beqn
&&\sigma_{non-diff}^{tagg}(dA\to nX)=
\int d^2b\int d^2r_T\,\left|\Psi_d(r_T)\right|^2\,
\exp\left[-\stn\,T_A^N\left(\vec b-
\vec r_T\right)\right]
\nonumber\\ &\times&
\left\{1-\exp\left[-\tilde\sinn\,T_A^N\left(b\right)
+4\,\seln\,T^N(\vec b-\vec r_T/2)\,\gamma(r_T)\,
\exp\left(-\frac{r_T^2}{4B_{NN}}\right)
\right]
\right\}
\label{42}
 \eeqn
 Here we made a correction for diffractive channels replacing
$\sinn\Rightarrow\tilde\sinn$ and $\seln\Rightarrow\tilde\seln$, valid
only for those experiment which are not sensitive to diffraction (PHENIX,
PHOBOS). Correspondingly, the numerical result for
$\sigma_{non-diff}^{tagg}(dA\to nX)$ is placed at the bottom part of the
Table. The results at the upper part of the Table use the total elastic
and total cross sections instead of $\tilde\seln$ and $\tilde\sinn$ as an
input for calculations. Indeed, since the STAR experiment is sensitive to
quasielastic nuclear excitation as well, it should be included, and one
has to rely on Eq.~(\ref{a.90}). We also demonstrate the impact parameter
dependence of $\sigma_{in}^{tagg}(dA\to nX)$ in Fig.~\ref{b-tagg-in}.
 \begin{figure}[tbh]
\includegraphics{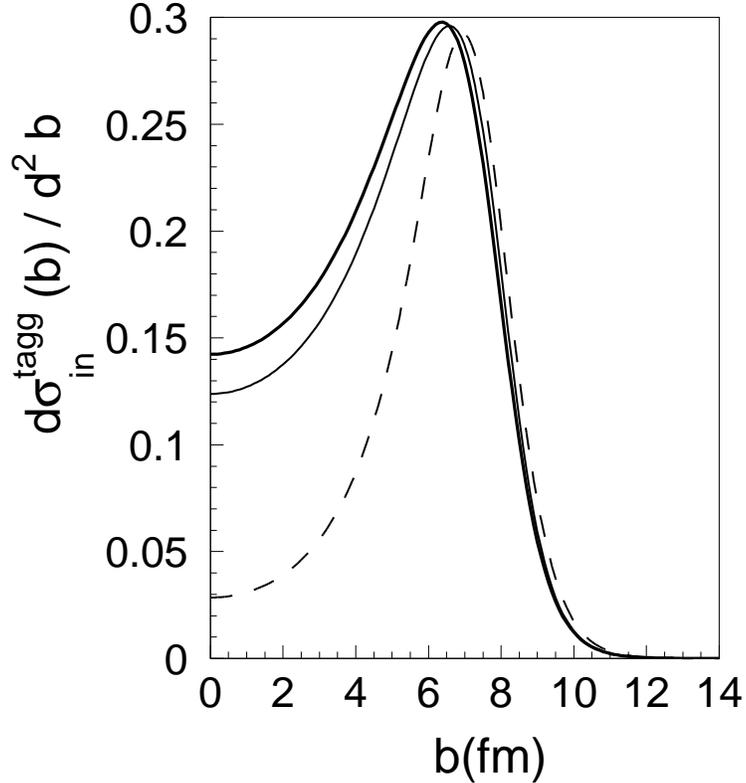}
 \begin{center}
 \vspace{10.5cm}
 \parbox{13cm} 
{\caption[Delta]
 {\sl The impact parameter distribution of interacting protons in tagged
deuteron-gold collisions with spectator neutrons. The calculation
includes diffractive excitations (STAR trigger).  Impact parameter $\vec
b$ corresponds to the proton. The dashed curve represents the Glauber
approximation Eq.~(\ref{39}). The thin solid curve include inelastic
shadowing related to excitation of the valence quark skeleton,
Eq.~(\ref{63}). The thick solid curve includes gluon shadowing as well.  
All curves are calculated with total cross section $\tilde\stn=51\mb$.}
 \label{b-tagg-in}} 
\end{center}
 \end{figure}
 Interesting that the interacting protons in tagged $dA$ collisions
strongly pick at the very edge of the nucleus in spite of the large radius
of the deuteron. This is not a trivial observation which can be probably
interpreted as follows. The spectator neutron must be mostly outside of
the nucleus. Then, for a protons which are close to the edge of the
nucleus the interval of azimuthal angle between the proton and neutron is
larger than for a proton deep inside the nuclear area. This phase space
factor enhances the contribution of peripheral protons.

To see how the spectator neutrons are distributed one can use the same
Eq.~(\ref{42}) with the replacement $\vec b \Rightarrow \vec b + \vec
r_T$. The result of calculations is depicted in Fig.~\ref{b-neutron-in}.
 \begin{figure}[tbh]
\includegraphics{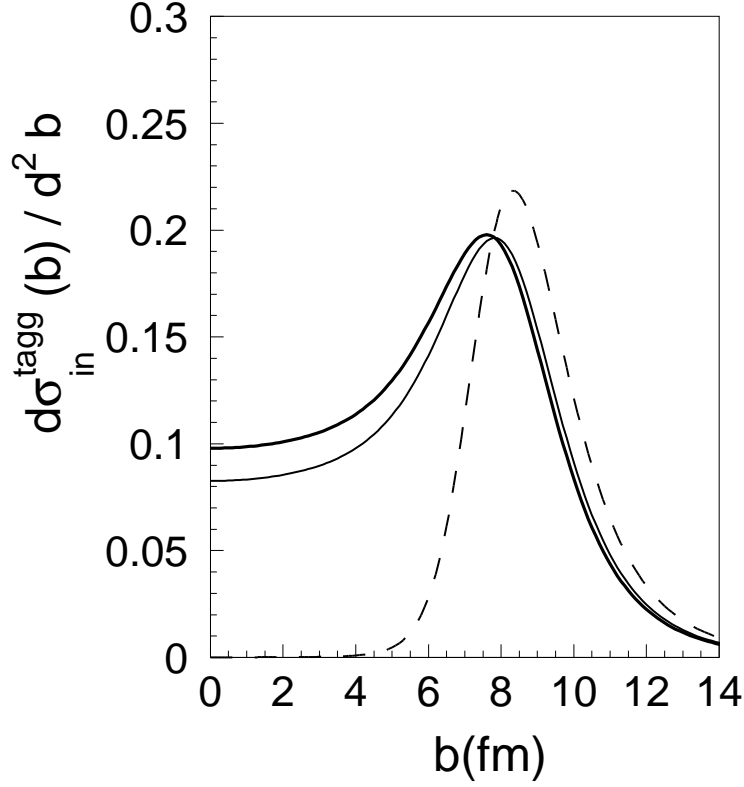}
 \begin{center}
 \vspace{10.5cm}
 \parbox{13cm} 
{\caption[Delta]
 {\sl The impact parameter distribution of spectator neutrons in tagged
deuteron-gold collisions with interacting protons. The calculation
includes diffractive excitations (STAR trigger).  Impact parameter $\vec
b$ corresponds to the proton. The dashed curve represents the Glauber
approximation Eq.~(\ref{39}). The thin solid curve include inelastic
shadowing related to excitation of the valence quark skeleton,
Eq.~(\ref{63}). The thick solid curve includes gluon shadowing as well.  
All curves are calculated with total cross section $\tilde\stn=51\mb$.}
 \label{b-neutron-in}} 
\end{center}
 \end{figure}
 This plot demonstrates that the spectators have a more peripheral impact
parameter distribution than the interacting protons, but they are
amazingly close.

We also calculated the number of collisions of the proton which
underwent interaction in events with a tagged spectator neutron,
 \beq
N_{coll}^{tagg} = \frac{\tilde\sinn}
{\sigma_{tagg}^{dA}}
\int d^2b\int d^2r_T\,\left|\Psi_d(r_T)\right|^2\,
T_A^N(\vec b + \vec r_T/2)\,
\exp\left[-\stn\,T_A^N(\vec b-
\vec r_T/2)\right]
\label{46}
 \eeq
 The results for $N_{coll}^{tagg}$ for events which include diffraction
or do not are shown at the upper and bottom panels of Table~\ref{tab}
respectively. The mean value of $N_{coll}$ for tagged events turns out to
be nearly a half of the minimal bias value, Eq.~(\ref{10}), which is for
two nucleons in the deuteron. This contradicts the intuitive expectation
that tagged events are much more peripheral than minimum bias inelastic
collisions and the proton should have a much smaller number of
collisions.

To get the transverse momentum distribution of spectator neutrons, one
should Fourier transform the elastic neutron amplitude before squaring
it,
 \beqn
\frac{d\sigma_{tagg}(dA\to nX)}{d^2q_T}
&=& \frac{1}{(2\pi)^2}\int d^2b
\left\{1-\exp\left[-\tilde\sinn\,T_A^N(b)
\right]\right\}
\int d^2r_1\,d^2r_2
\nonumber\\ &\times&
\exp\Bigl[i\vec q_T(\vec r_1-\vec r_2)\Bigr]
\exp\left\{-{1\over2}\,\stn\,\left[T_A^N\left(\vec b-
\vec r_1\right) + T_A^N\left(\vec b-\vec r_2\right)
\right]\right\}
\nonumber\\ &\times&
\int\limits_{-\infty}^\infty dr_L
\left[\frac{u^*(r_L,r_1)u(r_L,r_2)+
w^*(r_L,r_1)w(r_L,r_2)}
{\sqrt{(r_L^2+r_1^2)(r_L^2+r_2^2)}}
\right],
\label{47}
 \eeqn
 Important is that the proton inelastic interaction is incoherent, therefore we
should first sum up coherently all amplitudes of neutron elastic scattering for
the fixed impact parameter of the proton, then Fourier transform it, square and
after all integrate over the proton impact parameter. This is explicitly done
in (\ref{47}). Apparently, integration over $\vec q_T$ in (\ref{47}) leads to
the expression in Eq.~(\ref{42}). The $S$ and $D$ wave functions are presented
in Appendix~\ref{appendC}.

In Fig.~\ref{q-tagg-minb} we compare the normalized differential cross
section Eq.~(\ref{47}),
 \beq
R_{tagg}(q_T) = \frac{1}
{\sigma^{tagg}_{in}}\,
\frac{d\sigma_{tagg}^{dA}}{d^2q_T}\,
\label{47a}
 \eeq
 with the undisturbed primordial distribution of the neutron in the incoming
deuteron, also normalized to one,
 \beqn
\frac{dN^d_n}{d^2q}&=&\frac{1}{(2\pi)^2}
\int d^2r_1\,d^2r_2
\exp\Bigl[i\vec q_T(\vec r_1-\vec r_2)\Bigr]
\nonumber\\ &\times&
\int\limits_{-\infty}^\infty dr_L
\left[\frac{u^*(r_L,r_1)u(r_L,r_2)+
w^*(r_L,r_1)w(r_L,r_2)}
{\sqrt{(r_L^2+r_1^2)(r_L^2+r_2^2)}}
\right],
\label{47b}
 \eeqn
 \begin{figure}[tbh]
\includegraphics{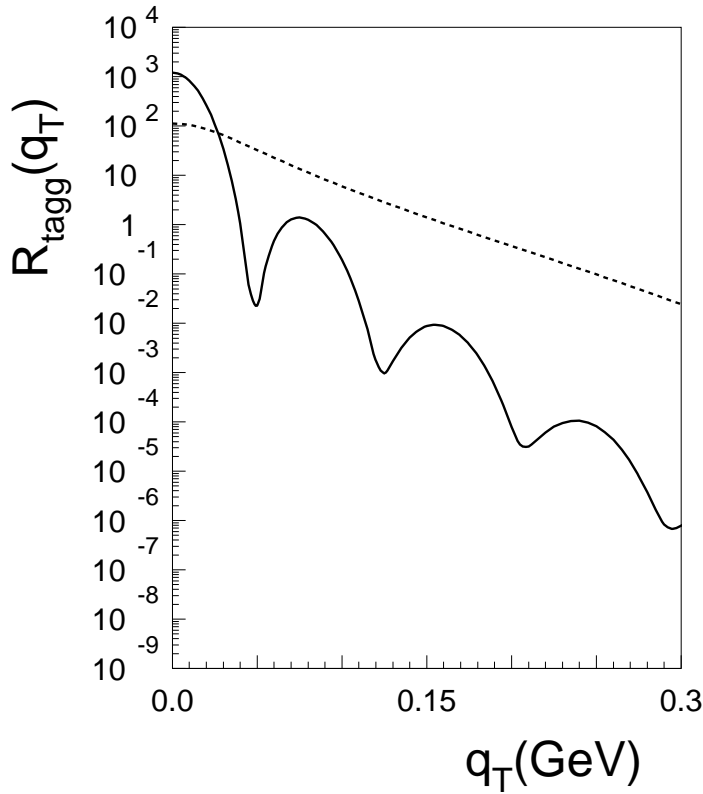}
 \begin{center}
 \vspace{10.5cm}
 \parbox{13cm} 
{\caption[Delta]
 {\sl Transverse momentum distribution of spectator neutrons
in the tagged reaction $d+Au\to n+X$ (solid curve), and
in the projectile deuteron (dashed curve). The inelastic reaction
$p+Au\to X$ is assumed to include diffraction (STAR experiment).
The calculations are performed in the Glauber approximation,
Eq.~(\ref{47}).}
 \label{q-tagg-minb}} 
\end{center}
 \end{figure}

 The surprising observation is that the spectator neutrons have a much
narrower $q_T$-distribution than the Fermi motion in the deuteron. This
is opposite to the usual $q_T$-broadening (Cronin effect) for particles
propagating though a matter \cite{jkt}. In the present case the nucleus
acts like a lens focusing neutrons. Fig.~\ref{q-tagg-minb} also exposes
quite a different shape of the $q_T$-distribution of spectators having
diffraction-like minima and maxima.

If to compare the mean values of $q_T^2$ of spectator neutrons with
the primordial value in the deuteron, the difference is tremendous,
about factor $20$.
 \beqn
\la q_T^2\ra_{spect} &=& 0.00038\GeV^2
\label{49a}\\
\la q_T^2\ra_{deuteron} &=& 0.0065\GeV^2\ 
\label{49b}
 \eeqn

This focusing effect is a beautiful manifestation of quantum mechanics.
The intuitive interpretation is rather straightforward. The condition that
the neutron in the deuteron remains intact, while the proton must
interact, means that the neutron tries to pass the nucleus through the
diluted periphery, while the proton prefers the collision to be central.  
These conflicting conditions cause a strong suppression of small-size
deuteron fluctuations, while large separations in the deuteron are
enhanced. Apparently, such large size configurations are related to a
smaller Fermi momentum and this simple observation explains the focusing
effect.

This explanation offers a possibility to study the correlation of the
focusing effect with centrality of collision\footnote{I am thankful to
Alexei Denisov for this suggestion.}. Suppressing $b$-integration in
(\ref{47}) one can trace the $b$-dependence of the focusing effect. In
Fig.~\ref{q-tagg-central} the same comparison of two $q_T$-distributions
is shown for central collision $b=0$ (impact parameter of the
interacting proton).
 \begin{figure}[tbh]
\includegraphics{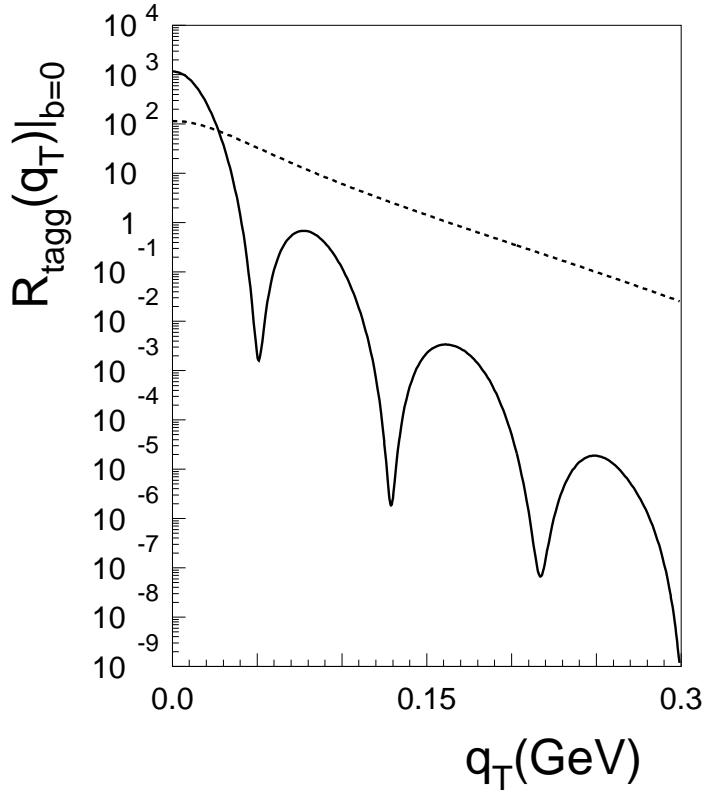}
 \begin{center}
 \vspace{10.5cm}
 \parbox{13cm} 
{\caption[Delta]
 {\sl Same as in Fig.~\ref{q-tagg-minb}, but for a central
($b=0$) proton-gold collision, accompanied by a spectator neutron.}
 \label{q-tagg-central}} 
\end{center}
 \end{figure}
The interpretation of central collisions is especially clear. Once the
proton hits the center of the nucleus, the spectator neutron must be
located along a rage ring outside of the nucleus with a radius larger
than the nucleus. Correspondingly, the $q_T$-distribution has the typical
diffractive shape and a small width $\Delta q_T \lsim 1/R_A$.

On the contrary to our expectations, the distributions are quite similar.
The mean value of $\la q_T^2\ra=0.00032\GeV^2$ is close to our result 
Eq.~(\ref{49a}) for the minimal bias sample. 

One may wonder why the minima on the $q_T$-distribution
Fig.~\ref{q-tagg-central} are deeper than for minimal bias sample
Fig.~\ref{q-tagg-minb}. In fact, for central collisions the minima go down to
zero, since we neglected the real part of the elastic amplitude and the
Fourier transform oscillates changing sign. However, the position of the
minima (slightly) depend on the impact parameter of the collision. Therefore,
when one sums up $q_T$-distributions with different minimum positions, the
resulting distribution will have minima which are partially filled up.

One should be cautious comparing these predictions with data which might be
contaminated by non-spectator neutrons. First, the neutron calorimeters 
used
at RHIC have a rather large acceptance which covers transverse momenta up to
$\sim 300\MeV$. Therefore, most of the neutrons which experienced quasielastic
scattering contribute as well (except STAR). Besides, the range of longitudinal
momenta is rather large and events with diffractive excitation on nucleons in
the gold should contribute too. All such neutrons are not spectators and have 
much wider $q_T$-distribution.

 Second, selecting central collisions in accordance with higher multiplicity,
one should remember that central collisions are suppressed (see
Fig.~\ref{b-tagg-in})  and one should not mix them up with the fluctuations of
multiplicity.

\section{Inelastic shadowing corrections}\label{in-corr}

\subsection{Intermediate state diffractive excitations}\label{sect-kk}

The Glauber model is a single-channel approximation, it misses the
possibility of diffractive excitation of the projectile in intermediate
state illustrated in Fig.~\ref{diff}.
 \begin{figure}[tbh]
\includegraphics{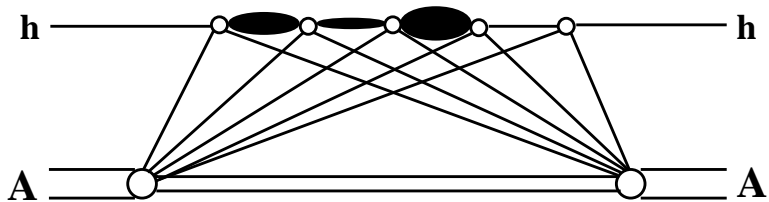}
 \begin{center}
 \vspace{4cm}
 \parbox{13cm} 
{\caption[Delta]
{\sl Diagonal and off-diagonal diffractive multiple interactions of the
projectile hadron in intermediate state.} 
\label{diff}} 
\end{center}
 \end{figure}
 These corrections called inelastic shadowing were introduced by
Gribov back in 1969 \cite{gribov}. The formula for the inelastic
corrections to the total hadron-nucleus cross section was suggested in
\cite{kk},
 \beqn
\Delta\sigma^{hA}_{tot} &=& 
- 4\pi\int d^2b
\exp\left[-{1\over2}\,\sigma^{hN}_{tot}\,
T_A(b)\right]
\int\limits_{M_{min}^2} dM^2\, 
\left.\frac{d\sigma_{sd}^{hN}}
{dM^2\,dp_T^2}\right|_{p_T=0}
\nonumber\\ &\times&
\int\limits_{-\infty}^{\infty}dz_1\,
\rho_A(b,z_1)
\int\limits_{z_1}^{\infty}dz_2\,
\rho_A(b,z_1)\,e^{iq_L(z_2-z_1)}\ ,
\label{54}
 \eeqn
 where $\sigma_{sd}^{hN}$ is the cross section of single diffractive 
dissociation $hN\to XN$ with longitudinal momentum transfer
 \beq
q_L=\frac{M^2-m_h^2}{2E_h}\ .
\label{56}
 \eeq

This correction makes nuclei more transparent \cite{kn}. One can also see
from Fig.~\ref{murthy} that (\ref{54}) does a good job describing data at
low energies \cite{murthy,gsponer}, since takes care of the onset of
inelastic shadowing via phase shifts controlled by $q_L$.  Higher order
off-diagonal transitions are neglected. Diagonal transitions (or
absorption of the excited state) are important, but unknown. Indeed, the
intermediate state $X$ has definite mass $M$, but no definite size, or
cross section. It is ad hoc fixed in (\ref{54}) at $\sigma_{tot}^{hN}$.
It has been a long standing problem how to deal simultaneously with phase
shifts which are controlled by the mass, and with the cross section which
depends on the size. This problem was eventually solved in
\cite{krt1,krt2} within the light-cone Green function approach (see
Sect.~\ref{gluons}).

The situation changes at the high energies of RHIC and LHC, all multiple
interactions become important, but phase shifts vanish, substantially
simplifying calculations. No experimental information, however, is
available for off-diagonal diffractive amplitudes for excited state
transitions $X_1\to X_2$. A solution proposed in \cite{kl} is presented
in the next Sect.~\ref{eigen}.

There is, however, one exclusion which is free of these problems,
hadron-deuteron collisions. In this case no interaction in intermediate
state is possible and knowledge of diffractive cross section $NN\to NX$
is sufficient to calculate the inelastic correction with no further
assumptions. In this case Eq.~(\ref{54}) takes the simple form
\cite{gribov,pr}, analogous to (\ref{a.58}),
 \beq
\Delta\sigma^{hd}_{tot} = - 
2\int dM^2\int dp_T^2\,
\frac{d\sdd}{dM^2dp_T^2}\,
F_d(t)\ .
\label{58}
 \eeq

We calculate this correction for $pd$ collisions following
\cite{anisovich} at $\sqrt{s}=200\GeV$ using the slope
$B_{NN}^{sd}=10\GeV^{-2}$ reduced compared to $B_{NN}^{el}=14\GeV^{-2}$
by $4\GeV^{-2}$ which is the proton vertex contribution to the elastic
slope. The upper cut off imposed by the deuteron formfactor on
integration over $M^2$ is quite high at this energy and we can use the
free diffraction cross section $\sigma_{sd}^{NN}=4\mb$ \cite{dino}. Then
we find $\Delta\sigma^{hd}_{tot} = - 1.75\mb$

\subsection{Eigenstate method}\label{eigen}

If a hadron were an eigenstate of interaction, i.e. could undergo only
elastic scattering (as a shadow of inelastic channels) and no diffractive
excitation was possible, the Glauber formula would be exact and no
inelastic shadowing corrections would be needed. This simple observation
gives a hint that one should switch from the basis of physical hadronic
states to a new one consisted of a complete set of mutually orthogonal
states which are eigenstates of the scattering amplitude operator. This
was the driving idea of description of diffraction in terms of elastic
amplitudes \cite{pom,gw}, and becomes a powerful tool for calculation of
inelastic shadowing corrections in all orders of multiple interactions
\cite{kl}. Hadronic states (including leptons, photons) can be decomposed
into a complete set of such eigenstates $|k\ra$,
 \beq
|h\ra=\sum\limits_{k}\,\Psi^h_k\,|k\ra\ , 
\label{b.1}
 \eeq 
 where $\Psi^h_k$ are hadronic wave functions in the form of Fock state
decomposition. They obey the orthogonality conditions,
 \beqn
\sum\limits_{k}\,\left(\Psi^{h'}_k\right)^{\dagger}\,\Psi^h_k
&=&\delta_{h\,h'}\ ;
\nonumber\\
\sum\limits_{h}\,\left(\Psi^{h}_l\right)^{\dagger}\,\Psi^h_k
&=&\delta_{lk}\ .
\label{b.2}
 \eeqn

We denote by $f_{el}^{kN}=i\,\sigma_{tot}^{kN}/2$ the eigenvalues of the
elastic amplitude operator $\hat f$ neglecting its real part. We assume
that the amplitude is integrated over impact parameter, {\it i.e.} that
the forward elastic amplitude is normalized as
$|f_{el}^{kN}|^2=4\,\pi\,d\sigma_{el}^{kN}/dt|_{t=0}$. We can then
express the elastic $f_{el}(hh)$ and off diagonal diffractive
$f_{sd}(hh')$ amplitudes as,
 \beq
f_{el}^{hN}=2i\,\sum\limits_k\,\left|
\Psi^h_k\right|^2\,\sigma_{tot}^{kN}
\equiv 2i\,\la\sigma\ra\ ;
\label{b.3}
 \eeq
 \beq
f_{sd}^{hN}(h\to h')=
2i\,\sum\limits_k\, (\Psi^{h'}_k)^{\dagger}\,
\Psi^h_k\,\sigma_{tot}^{kN}\ . 
\label{b.4}
 \eeq
 Note that if all the eigen amplitudes were equal, the diffractive
amplitude (\ref{b.4}) would vanish due to the orthogonality relation,
(\ref{b.2}). The physical reason is obvious. If all the $f_{el}^{kN}$ are
equal, the interaction does not affect the coherence between the
different eigen components $|k\ra$ of the projectile hadron $|h\ra$.
Therefore, off diagonal transitions are possible only due to differences
between the eigen amplitudes.

If one sums up all final states in the diffractive cross section
one can use the completeness condition (\ref{b.2}).
Then, excluding the elastic channels one gets \cite{kl,mp,zkl},
 \beq
16\pi\,\frac{d\sigma^{hN}_{sd}}{dt}\biggr|_{t=0}=
\sum\limits_i \left|\Psi^h_i\right|^2
\left(\sigma^{iN}_{tot}\right)^2
- \biggl(\sum\limits_i 
\left|\Psi^h_i\right|^2\sigma^{iN}_{tot}\biggr)^2
\equiv \la\sigma_{tot}^2\ra - \la\sigma_{tot}\ra^2\ .
\label{b.6}
 \eeq

 As far as the main problem of Glauber approximation is the need to
include off-diagonal transitions, one should switch to an eigenstate
basis. Then each of the eigen states can experience only elastic
diffractive scatterings and the Glauber eikonal approximation becomes
exact. Thus, all expressions for cross sections of different channels
derived in Glauber approximation in Appendix~\ref{appendA} are exact for
any of the eigenstates. Then, the corresponding cross sections for
hadron-nucleus collisions are obtained via a proper averaging of those in
Appendix~A \cite{kl,zkl},
 \beq
\sigma^{hA}_{tot} = 2\int d^2b\,\left\{1 -
\left\la\exp\left[-{1\over2}\,\sigma_{tot}\,T^h_A(b)\right]
\right\ra\right\}
\label{b.10}
 \eeq
 \beq
\sigma^{hA}_{el} = \int d^2b\,\left|1 -
\left\la\exp\left[-{1\over2}\,\sigma_{tot}\,T^h_A(b)\right]
\right\ra\right|^2
\label{b.15}
 \eeq
 \beq
\sigma^{hA}_{in} = \int d^2b\,\left\{1 -
\left\la\exp\Bigl[-\sigma_{in}\,T^h_A(b)\Bigr]
\right\ra\right\}
\label{b.20}
 \eeq

It is interesting that the last expression for $\sina$ is already free of
diffraction contribution. Although only elastic and quasi-elastic cross
sections were subtracted from $\sta$ in Glauber model in Appendix~A,
after averaging over eigenstates it turns out that diffraction is
subtracted as well. Indeed, direct averaging of the elastic cross
section Eq.~(\ref{a.60}) is different from (\ref{b.15}) and includes
coherent diffraction, $hA\to XA$, which cross section reads 
\cite{kl,zkl},
 \beq
\sigma^{hA}_{sd}(hA\to XA) = \int d^2b\,\left\{
\Bigl\la\exp\left[-\sigma_{tot}\,T^h_A(b)\right]
\Bigr\ra\ -
\left\la\exp\left[-{1\over2}\,\sigma_{tot}\,T^h_A(b)\right]
\right\ra^2\right\} 
\label{b.25}
 \eeq
 Averaging of the quasi-elastic cross section Eq.~(\ref{a.90}) leads to
inclusion of diffractive excitation of the hadron $h\to X$ besides
excitation of the nucleus, $A\to Y$.

Thus, Eq.~(\ref{b.20})  resulting from a direct averaging of the single
channel inelastic cross section Eq.~(\ref{a.100})  corresponds to the
part of the total $hA$ cross section which does not contain, elastic
scattering, $hA\to hA$, coherent diffraction, $hA\to XA$, quasi-elastic,
$hA\to hY$, and double diffraction, $hA\to XY$. This part of the cross
section is what is measured as the inelastic cross section in heavy ion
and $p(d)A$ collisions at SPS and RHIC, and what we are going to
calculate below.

One may wonder, what is the difference between the cross sections
Eqs.~(\ref{b.10})-(\ref{b.20}) and those in Glauber approximation,
Eqs.~(\ref{a.50}), (\ref{a.60}) and (\ref{a.100})? The difference is
obvious, in the former set of equations the exponentials are averaged,
while in the Glauber approximation contains exponentials of averaged
values. For instance, the total cross section in the Glauber 
approximation reads,
 \beq
\sigma^{hA}_{tot}\Bigr|_{Gl} = 2\int d^2b\,\left\{1 -
\exp\left[-{1\over2}\,\la\sigma_{tot}^i\ra\, T^h_A(b)\right]
\right\}\ ,
\label{b.30}
 \eeq
 where $\la\sigma_{tot}\ra = \st$. If to subtract this from
Eq.~(\ref{b.10}), the rest is the Gribov's inelastic correction
calculated in all orders. Indeed, we can compare it with the expression
Eq.~(\ref{b.10}) expanding the exponentials in (\ref{b.10}) and
(\ref{b.30}) in multiplicity of interactions up to the lowest order.
Employing (\ref{b.6}) we find,
 \beq
\sigma^{hA}_{tot} - \sigma^{hA}_{tot}\Bigr|_{Gl} =
\int d^2b\, {1\over4}\,\Bigl[\la\sigma^i_{tot}\ra^2 -
\la(\sigma^i_{tot})^2\ra\Bigr]\,T^h_A(b)^2
=- 4\pi\int d^2b\,T^h_A(b)^2\int dM^2\,
\frac{d\sigma^h_{sd}}{dM^2dt}\biggr|_{t=0}\ .
\label{b.40}
 \eeq
 This result is identical to Eq.~(\ref{54}), if to neglect there the 
phase shift vanishing at high energies, and also to expand the 
exponential.

Note that since the inelastic nuclear cross section in the form
Eq.~(\ref{a.100}) is correct for eigenstates, one may think that
averaging this expression would give the correct answer. However, such a
procedure includes possibility of excitation of the projectile and
disintegration of the nucleus to nucleons, but misses possibility of
diffractive excitation of bound nucleons which is not a small correction.
We introduce a corresponding correction in the next section.

\section{Light-cone dipoles and inelastic shadowing}
\label{dipole}
\subsection{Excitation of the valence quark skeleton}
\label{quarks}

The light-cone dipole representation in QCD was introduced in \cite{zkl}
where it was realized that color dipoles are the eigenstates of
interaction and can be an effective tool for calculation of diffraction
and nuclear shadowing.  It was concluded that the key quantity of the
approach, the cross section of the dipole-nucleon, $\sq(r_T)$, is
a universal and flavor independent function which depends only on
transverse
separation $r_T$ and energy. Of course the energy must be sufficiently
high to freeze variations of the dipole size during interaction,
otherwise one should rely on the Green function approach \cite{kz91,krt1,
krt2} (see Sect.~\ref{gluons}).

This representation suggests an effective way to sum up all multi-step
inelastic corrections in all orders \cite{zkl}. Since dipoles are
eigenstates of interaction in QCD, they are not subject to any
diffractive excitation, and the eikonal approximation becomes exact.  
Therefore, if energy is high enough to keep the transverse size of a
dipole "frozen" by Lorentz time dilation during propagation through the
nucleus, one can write the cross sections in the form
Eqs.~(\ref{b.10})-(\ref{b.20}). The averaging in this case means
summing-up different Fock components of the hadron consisted of different
numbers of quarks and gluons, and for each of them integration over $r_T$
(intrinsic separations), weighted with the square of the hadron
light-cone wave function $|\Psi_h(r_T)|^2$. We assume that the hadron
does not have a ''molecular'' structure, i.e. is not like a deuteron
consisting of two colorless clusters. Therefore all following expressions
apply only to elementary hadrons. To simplify calculations, in what
follows we rely on the quark-diquark model of the proton, neglecting the
diquark size. The total cross section is basically insensitive to the
diquark size, besides, there are many evidences that this size is indeed
small \cite{diquark,kz-ann}.

\subsubsection{Nuclear transparency}\label{transparency}

According to the Glauber model hadrons attenuate exponentially in nuclear matter,
 \beq
Tr = \exp(-\st\,T_A)\ ,
\label{52}
 \eeq
 where $Tr$, called nuclear transparency, is the survival
probability of a hadron propagating through a nuclear matter of thickness
$T_A$. However, we know that the hadron fluctuates and can be viewed as a
combination of Fock states of different content and size. Some of them
having a small transverse size can easily penetrate the medium and do not
attenuate as fast as in (\ref{52}). 

Assuming that the hadronic wave function has a Gaussian form and the
dipole cross section $\sigma(r_T)\propto r_T^2$ (this small-$r_T$
behavior does a good job describing hierarchy of hadronic cross sections
and their sizes \cite{ph}) we can perform averaging in
(\ref{b.10}) and arrive at a rather simple expression \cite{zkl},

 \beq
\Biggl\la\exp[-\sigma(r)\,T_A]\Biggr\ra =
\frac{1}{1+\st\,T_A}\ .
\label{55}
 \eeq
 This explicitly demonstrates how Gribov's corrections make nuclei more
transparent. Since exponential attenuation is much stronger than a power,
for large $T_A$ (central collisions with heavy nuclei) the difference
might be tremendous.

\subsubsection{Cross sections}\label{Xsection}.

{\bf The total cross section}.

 The total hadron-nucleus cross section is modified according to
(\ref{55}) as,
 \beq
\sta = \int d^2b\ 
\frac{\st\,T^h_A(b)} 
{1+{1\over2}\st\,T^h_A(b)}\ ;
\label{60}
 \eeq

Although the Gribov's corrections (color transparency) make nuclei much
more transparent, the modified total cross section Eq.~(\ref{60}) is not
much smaller than the result of Glauber approximation Eq.~(\ref{a.50}).
This is because the central area of a heavy nucleus is "black",
i.e. fully absorptive, in both cases, and the cross section is mainly
related to the geometry of the nucleus. In other words, the exponential
term in (\ref{a.50}) is very small for central collisions, and the total
cross section is rather insensitive to even dramatic variations of its
magnitude. 

{\bf The elastic cross section} 

The partial elastic cross section is given by the square of the averaged
value of the elastic amplitude. We get,
 \beq
\sela = {1\over4}\int d^2b\ 
\frac{[\st\,T^h_A(b)]^2}
{[1+{1\over2}\st\,T^h_A(b)]^2}\ .
\label{62}
 \eeq

Correspondingly, the differential elastic cross section reads,
 \beq
\frac{d\sela}{dq_T^2} = {1\over16\pi}\left|\int d^2b\ 
\frac{\st\,T^h_A(b)}
{1+{1\over2}\st\,T^h_A(b)}\,
\exp(i\vec q\cdot\vec b)\right|^2\ .
\label{62a}
 \eeq

{\bf The total inelastic cross section} 

The cross section of all inelastic channels is given by the difference,
 \beq 
\sina= \sta-\sela = \int d^2b\ 
\frac{\st\,T^h_A(b)\left[1+{1\over4}\st\,T^h_A(b)\right]}
{\left[1+{1\over2}\st\,T^h_A(b)\right]^2}\ .
\label{63}
 \eeq
  This cross section approaches the unitarity limit for $\st\,T^h_A(b)\gg
1$ at the nuclear center, but is proportional to $T^h_A(b)$ at the
nuclear periphery.

{\bf Diffractive excitation of the hadron}. 

The combined cross section of elastic scattering and diffraction when the
hadron may be either excited or not, but the nucleus remains intact, is
given by the average of the $dA$ elastic partial amplitude squared,
 \beq
\sigma^{hA}_{sd+el}(hA\to XA) = {1\over2}\int d^2b\ 
\frac{[\st\,T^h_A(b)]^2}
{[1+\st\,T^h_A(b)][1+{1\over2}\st\,T^h_A(b)]}\ .
\label{64}
 \eeq
 Here we firstly averaged over the quark coordinates in the nucleons,
secondly, squared the result, and thirdly, subtracted the elastic $dA$
cross section [compare with (\ref{b.25})].

{\bf Diffractive excitation of the nucleus}. 

The cross section of the reaction where the nucleus is diffractively
excited, and the hadron either remains intact or is excited too,
reads,
 \beq
\sigma^{hA}_{qel}(hA\to XA^*) = \int d^2b\ 
\frac{2\,\tilde\sel\, T^h_A(b)}
{[1+\st\,T^h_A(b)]^3}\ ,
\label{66}
 \eeq
 where 
\beq
\tilde\sel = \sel+\sdd(hN\to XN))+
\sdd(hN\to hY)) + \sddd(hN\to XY))\ ,
\label{68}
 \eeq
 and $\sdd$ is a cross section of single diffractive excitation of either
the beam or the target; the double diffractive cross section $\sddd$
corresponds to diffractive excitation of both.

 Deriving Eq.~(\ref{66})  we made use of smallness of the elastic cross
section and expanded the exponential. Higher orders of $\sel$ are
neglected, but the corrections are easy to calculate. We also neglected
the small variation of the elastic slope of the dipole-nucleon cross
section with $r_T$.

Eq.~(\ref{66}), as one can see from (\ref{68}), takes into account
possibility of diffractive excitation of the projectile. This is a direct
consequence of the eigenstate approach. In addition, we also included the
possibility of diffractive excitation of bound nucleons in the 
target.  Those excitations are not shadowed by
multiple interactions in the nucleus, since all extra particles produced
this way stay in the nuclear fragmentation region and do not break down
the large rapidity gap structure of the event. Therefore, they may be
incorporated into $\tilde\sel$ adding the two last terms.
At $\sqrt{s}=200\GeV$ single and double diffraction cross
section are about equal, $\ssdn\approx\sddn\approx 4\mb$ \cite{dino,cdf},
$\sigma^{NN}_{el}\approx 9\mb$, so $\tilde\sigma^{NN}_{el}\approx 21\mb$.

Diffractive reactions Eq.~(\ref{66})-(\ref{68}) do not produce any 
particles at central rapidities. Therefore, if one wants to calculate
the part of the total hadron-nucleus cross section detected 
experimentally, one should subtract these diffractive contributions,
 \beqn
\tilde\sina&=&\sta-\sigma^{hA}_{sd+el}(hA\to XA)-
\sigma^{hA}_{qel}(hA\to XA^*)
\nonumber\\ &=&
\int d^2b\ 
\frac{\st\,T^h_A(b)}
{1+\st\, T^h_A(b)} \left\{1-
\frac{2\,\tilde\sel/\st}  
{[1+\st\,T^h_A(b)]^2}\right\}\ .
\label{72}
 \eeqn
 Since $pp$ cross section is used as a baseline for comparison, the same 
subtraction should be done in this case too,
 \beq
\tilde\sigma^{pp}_{in}= 
\sigma^{NN}_{tot} - \tilde\sigma^{NN}_{el}\ ,
\label{80}
 \eeq
 what comes to about $\tilde\sigma^{pp}_{in}=30\mb$ at 
$\sqrt{s}=200\GeV$.

Then, the number of collisions at given impact parameter corrected for 
inelastic shadowing reads,
 \beq
N_{coll}(b) = 
\frac{\tilde\sigma^{NN}_{in}}{\sigma^{NN}_{tot}}
\Bigl[1+\stn\,T^h_A(b)\Bigr]
\left\{1-\frac{\tilde\seln/\stn}
{[1+\stn\, T^h_A(b)]^2}\right\}^{-1}\ .
\label{90}
 \eeq

\subsection{Deuteron-nucleus collisions}\label{dA} 

So far we considered the case of a colorless hadrons, but colored
constituents. The specifics of a deuteron is that it contains two
colorless clusters, nucleons.  Therefore, one of the inelastic
corrections which we already took into account in (\ref{38})  is related
to fluctuations of the deuteron size. The next step is to average over
the fluctuations of the sizes of the nucleons.

{\bf The total deuteron-nucleus cross section}.\\  Now we should average
$\sigma^{dA}_{tot}$ over the inter-nucleon separation, as well as over the
nucleon sizes, $\vec r_1$ and $\vec r_2$,
 \beq
\sigma^{dA}_{tot} = 2\int d^2b\int d^2r_T\,
\Bigl|\Psi_d(r_T)\Bigr|^2\,
\left\la f^{dA}(\vec b,\vec r_T)
\right\ra_{r_1,r_2}\ ,
\label{92}
 \eeq
 where
 \beqn
\left\la f^{dA}(\vec b,\vec r_T)
\right\ra_{r_1,r_2} &=&
1 - \frac{1}{[1+{1\over2}\,\stn\,
T^N_A(\vec b+{1\over2}\,\vec r_T)]
[1+{1\over2}\,\stn\,T^N_A(\vec b-{1\over2}\,\vec r_T)]} 
\nonumber\\ &-&
\frac{\seln\,T_A^N(b)\,\exp\left(-\frac{r_T^2}{4B_{NN}}\right)}
{[1+{1\over2}\,\stn\,T^N_A(\vec b+{1\over2}\,\vec r_T)]^2
[1+{1\over2}\,\stn\,T^N_A(\vec b-{1\over2}\,\vec r_T)]^2}\ .
\label{94}
 \eeqn

The result of calculation exposed in Table~\ref{tab} is smaller than the
Glauber model value. The difference comes from inelastic shadowing
related to diffractive excitations of the colorless clusters in the
deuteron, each consisted of three valence quarks.

{\bf Elastic and diffractive scattering of deuterons}.\\
Correspondingly, the total cross section of elastic scattering and
diffractive excitation of the deuteron has the form,
 \beq
\sigma^{dA}_{el} +\sigma^{dA}_{sd}(dA\to XA) = 
\int d^2b\int d^2r_T\,
\Bigl|\Psi_d(r_T)\Bigr|^2\,
\left\la g^{dA}(\vec b,\vec r_T)
\right\ra_{r_1,r_2}\ ,
\label{96}
 \eeq
 where
 \beqn
\left\la g^{dA}(\vec b,\vec r_T)
\right\ra_{r_1,r_2} &=&
1 - \frac{2}{[1+{1\over2}\,\stn\,
T^N_A(\vec b+{1\over2}\,\vec r_T)]
[1+{1\over2}\,\stn\,T^N_A(\vec b-{1\over2}\,\vec r_T)]} 
\nonumber\\ &+&
\frac{2\,\seln\,T_A^N(b)\,
\exp\left(-\frac{r_T^2}{4B_{NN}}\right)}
{[1+\stn\,T^N_A(\vec + \vec r_T)]^2
[1+\stn\,T^N_A(\vec b-{1\over2}\,\vec r_T)]^2}
\nonumber\\ &+&
\frac{1}{[1+\stn\,
T^N_A(\vec b+\vec r_T)]
[1+{1\over2}\,\stn\,T^N_A(\vec b-{1\over2}\,\vec r_T)]} 
\nonumber\\ &-&
\frac{2\,\seln\,T_A^N(b)\,
\exp\left(-\frac{r_T^2}{4B_{NN}}\right)}
{[1+\stn\,T^N_A(\vec b+{1\over2}\,\vec r_T)]^2
[1+\stn\,T^N_A(\vec b-{1\over2}\,\vec r_T)]^2}\ .
\label{120}
 \eeqn

{\bf Inelastic deuteron-nucleus collisions}.\\ If to subtract the elastic
and diffractive cross section Eq.~(\ref{96}) from (\ref{92}) the rest
will be the inelastic cross section which covers all diffractive
excitations of the nucleus, but not gold. This is what is measured at the
STAR experiment. To comply with the condition of experiments insensitive
to diffraction one should also subtract the cross section of diffractive
excitation of the nucleus. The results read [compare with (\ref{72})],
 \beqn
\tilde\sigma^{dA}_{in} &=&
\sigma^{dA}_{tot}-\sigma^{dA}_{el}-\sigma^{dA}_{sd}(dA\to XA)
- \sigma^{dA}_{sd}(dA\to dY) - \sigma^{dA}_{dd}(dA\to XY) 
\nonumber\\ &=&
\int d^2b\int d^2r_T\,
\Bigl|\Psi_d(r_T)\Bigr|^2\,
\left\la h^{dA}(\vec b,\vec r_T)
\right\ra_{r_1,r_2}\ ,
\label{124}
 \eeqn
 where
\beqn
\left\la h^{dA}(\vec b,\vec r_T)
\right\ra_{r_1,r_2} &=&
\left\{1-\frac{1}{[1+\stn\,
T^N_A(\vec b+{1\over2}\vec r_T)]
[1+\stn\,T^N_A(\vec b-{1\over2}\,\vec r_T)]} 
\right.
\nonumber\\ &-& \left. 
\frac{2\,\tilde\seln\,
T^N_A(\vec b+{1\over2}\,\vec r_T)}
{[1+\stn\,T^N_A(\vec b+{1\over2}\,\vec r_T)]^3
[1+\stn\,T^N_A(\vec b-{1\over2}\,\vec r_T)]}
\right.\nonumber\\ &-& \left. 
\frac{2\,\tilde\seln\,
T^N_A(\vec b-{1\over2}\,\vec r_T)}
{[1+\stn\,T^N_A(\vec b+{1\over2}\,\vec r_T)]
[1+\stn\,T^N_A(\vec b-{1\over2}\,\vec r_T)]^3}
\right.\nonumber\\ &-& \left. 
\frac{2\,\seln\,T_A^N(b)\,
\exp\left(-\frac{r_T^2}{4B_{NN}}\right)}
{[1+\stn\,T^N_A(\vec b+{1\over2}\,\vec r_T)]^2
[1+\stn\,T^N_A(\vec b-{1\over2}\,\vec r_T)]^2}
\right\}\ .
\label{128}
 \eeqn

The results of calculations of both inelastic cross sections with and
without nuclear diffraction, as well as the corresponding numbers
collisions which are rather small compared to what was calculated in
\cite{phenix}, are presented in Table~\ref{tab}. As expected, the cross
sections are smaller than predicted by the Glauber model, while the
numbers of collisions are larger. We also plotted $b$-dependence of
$\sigma^{dA}_{in}$ in Fig.~\ref{results-in} (thin solid curve). Comparing
with the Glauber curve we see that this class of inelastic shadowing
corrections leave the mid of nucleus ``black'', but makes it rather
transparent on the periphery.

{\bf Production of spectator nucleons}.\\
Similarly, one derives an equation for the cross section of a channel
with tagged spectator nucleons corrected for inelastic shadowing,
 \beqn
\sigma^{dAu}_{tagg} &=& 
\int d^2b\int d^2r_T\,
\frac{\Bigl|\Psi_d(r_T)\Bigr|^2}
{1+\stn\,T_A^N(\vec b+{1\over2}\vec r_T)}
\left\{1 - \frac{1}{1+\stn\,T_A^N(\vec b-{1\over2}\vec r_T)}
\right.\nonumber\\ &-& \left.
\frac{2\,\tilde\seln\,T_A^N(\vec b-{1\over2}\vec r_T)
+2\,\seln\,T_A^N(b)\,\exp[-r_T^2/4B_{NN}]}
{\left[1+\stn\,T_A^N(\vec b+{1\over2}\vec r_T)\right]^3}
\right\}.
\label{130}
 \eeqn

Events with tagged nucleons are especially sensitive to the transparency
of the nucleus.  We calculated the cross section, Eq.~(\ref{130}), and
the results, as well as the corresponding numbers of collisions are shown
in Table~\ref{tab}.  The effect of inelastic corrections on the impact
parameter distribution of interacting protons in tagged events with a
spectator neutron is demonstrated in Fig.~\ref{b-tagg-in}. Calculation was
done for inelastic proton interaction including diffractive excitations
(STAR). As one could anticipate, the nucleus becomes much more
transparent in the center. Indeed, for a nearly black nucleus inelastic
corrections keep it black since transparency, or the exponential term is
so small that even if it is modified by a large factor, the final change
is very small. However, tagged events is a direct measure of
transparency, and the inelastic corrections are maximal in this case. It
is not surprising that $N_{coll}$ is quite large (considering that only
one nucleon interacts.

\subsection{Towards realistic calculations}\label{real}

\subsubsection{Three valence quarks}.

For the sake of simplicity we used so far the approximation of a
quark-diquark structure of the proton, and neglected the diquark size. As
long as the diquark is indeed as small as $0.2-0.3\fm$
\cite{diquark,kz-ann}, this approximation is rather precise even for
heavy nuclei which hardly can resolve such a small size. However, the
mean size of the isoscalar diquark is still a debatable issue, besides,
an isovector diquark is probably a big object. Then, one may expect
nuclear matter to be more opaque for a high energy nucleon compared to
what was found above.

We evaluate nuclear transparency for another extreme, i.e. for the case
of a proton wave function symmetric in all quark coordinates, with a mean
size of any diquark of the order of $0.7\fm$,
 \beq
\left|\Psi_N(\vec r_1,\vec r_2,\vec r_3)\right|^2 = 
\frac{3}{(\pi\,r_N^2)^2}
\exp\left(-\frac{r_1^2+r_2^2+r_3^2}{r_N^2}\right)\,
\delta(\vec r_1+\vec r_2+\vec r_3)\ .
\label{135}
 \eeq

To perform the averaging of the eikonal exponentials in
(\ref{b.10})-(\ref{b.20}) we need to know the three-body  dipole cross
section, which we express via the conventional $\bar qq$ one as,
 \beq
\sigma_{3q}(\vec r_1,\vec r_2,\vec r_3) =
{1\over2}\,\Bigl[\sigma_{\bar qq}(r_1)+
\sigma_{\bar qq}(r_2)+
\sigma_{\bar qq}(r_3)\Bigr]\ .
\label{140}
 \eeq
 It satisfies the limiting conditions, namely, turns into $\sigma_{\bar
qq}(r)$ if one of three separations is zero. Assuming that $\sigma_{\bar
qq}(r)=Cr^2$, this cross section averaged with the wave function squared
Eq.~(\ref{135}) gives $\stn=Cr_N^2/2$.

Now we can calculate the nuclear transparency averaging the eikonal
exponential,
 \beqn
\Bigl\la\exp\left[- \sigma_{3q}(r_i)T_A(b)\right]\Bigr\ra &=&
\int \prod\limits_i^3 d^2r_i\,
\left|\Psi_N(\vec r_1,\vec r_2,\vec r_3)\right|^2\,
\exp\left[-\sigma_{3q}(\vec r_1,\vec r_2,\vec r_3)\,
T_A(b)\right]
\nonumber\\ &=&
\frac{1}{\left[1+{1\over2}\,
\stn\,T_A(b)\right]^2}
\label{145}
 \eeqn
 We see that nuclear transparency in this case is quadratic, rather than
linear function of the inverse nuclear thickness. For small $\stn\,T_A(b)
\ll 1$ it coincides with the result of the quark-diquark model,
Eq.~(\ref{55}), however falls steeper at large $T_A(b)$. This is not
surprising: in order to make use of color transparency the whole proton
has to fluctuate into a small transverse area, and it is more probable
for a two- than three-body system.

One can consider these results as a lower [Eq.~(\ref{145})] and an upper
[Eq.~(\ref{55})] bound for nuclear transparency.  We calculated different
cross sections using the average of the eikonal exponential in the form
Eq.~(\ref{145}) instead of Eq.~(\ref{55}), and the results are shown in
Table~\ref{tab} in parenthesis. Unfortunately, we still do not know the
proton wave function sufficiently well to fix this uncertainty for
nuclear transparency. Nevertheless, the difference is not large for
real nuclei. For instance, the inelastic non-diffractive $d-Au$ cross
section presented in Table~\ref{tab} increases by about $6\%$.

\subsubsection{Realistic dipole cross section}.

The dipole cross section $\sq\propto r_T^2$ used above is justified only
for small $r_T$, while it is expected to level off at large $\bar qq$
separations. More reliable calculations can be done using a realistic
phenomenological cross section. A quite popular parametrization was
proposed in \cite{gbw} and fitted to HERA data for $F_2(x,Q^2)$. However,
it should not be used for our purpose, since is unable to provide the
correct energy dependence of hadronic cross sections. Namely, the
pion-proton cross section cannot exceed $23\mb$ \footnote{According to
\cite{sgbk} this dipole cross section reproduced well the energy
dependence of the photoabsorption cross section $\sigma^{\gamma
p}_{tot}(s)$. This happens only due to the singularity in the light-cone
wave function of the photon at small $r_T$. This is a specific property
of the transverse photon wave function and is not applicable to
hadrons.}.

A parametrization more appropriate for soft hadronic physics was proposed
in \cite{kst2}:
 \beq
\sigma_{\bar qq}(r_T,s)=\sigma_0(s)\,\left[
1-{\rm exp}\left(-\frac{r_T^2}
{R_0^2(s)}\right)\right]\ ,
\label{170}
 \eeq
 where $R_0(s)=0.88\,fm\,(s_0/s)^{0.14}$ and $s_0=1000\,GeV^2$. In
contrast to \cite{gbw} all values depend on energy (as it is supposed to
be for soft interactions) rather than on $x$ and the energy
dependent parameter $\sigma_0(s)$ is defined as,
 \beq
\sigma_0(s)=\sigma^{\pi p}_{tot}(s)\,
\left(1 + \frac{3\,r^2_0(s)}{8\,\la r^2_{ch}\ra_{\pi}}
\right)\ ,
\label{180}
 \eeq
 Here $\la r^2_{ch}\ra_{\pi}=0.44\pm 0.01\,fm^2$ \cite{pion} is the mean
square of the pion charge radius. Cross section (\ref{170}) averaged
with the pion wave function squared automatically reproduces the
pion-proton cross section. The $pp$ total cross section is also well
reproduced using the quark-diquark approximation for the proton wave
function. The parameters are adjusted to HERA data for the proton
structure function. Agreement is quite good up to at least $Q^2\sim
10\,GeV^2$ sufficient for our purposes.

With such a dipole cross section one can perform analytic calculations
expanding the Glauber exponentials in (\ref{b.10})-(\ref{b.25}). Then the
total cross section gets the form,
 \beq
\sta=2\int d^2b\,\left\{1-
\exp\left[-{1\over2}\sigma_0(s)T^h_A(b)\right]
\sum\limits_{n=0}^\infty
\frac{[\sigma_0(s)T^h_A(b)]^n}
{2^n\,n!\,(1+n\,\delta)}\,
\right\}\ ;
\label{200}
 \eeq

Correspondingly, the sum of elastic and diffractive deuteron scattering
on the nucleus reads,
 \beqn
\sigma^{hA}_{sd+el}(hA\to XA) &=&
\int d^2b\
\Biggl\{1+\exp\left[-{1\over2}
\sigma_0(s)T^h_A(b)\right]
\nonumber\\ &\times& \left.
\sum\limits_{n=0}^\infty
\frac{[\sigma_0(s)T^h_A(b)]^n}
{n!\,(1+n\,\delta)}\ 
\left(1-2^{1-n}\exp\left[-{1\over2}
\sigma_0(s)T^h_A(b)\right]\right)
\right\}\ .
\label{210}
 \eeqn

The cross section of quasielastic excitation of the nucleus with
simultaneous possibility to excite the deuteron is given by, 
 \beqn
\sigma^{hA}_{qel}(hA\to XA^*) &=&
\int d^2b\
\exp\left[-{1\over2}
\sigma_0(s)T^h_A(b)\right]\,
\nonumber\\ &\times& 
\sum\limits_{n=0}^\infty
\frac{[\sigma_0(s)\,T^h_A(b)]^n}{n!}\ 
\frac{2\delta^2}
{[1+n\delta][1+(n+1)\delta][1+(n+2)\delta]}\ .
\label{220}
 \eeqn

In all these equations
 \beq
\delta=\frac{8\la r_p^2\ra}{3R_0^2(s)}\ .
\label{230}
 \eeq

Now one can calculate $\tilde\sina$ subtracting (\ref{210}) and
(\ref{220}) from (\ref{200}). However, in this paper we restrict
ourselves by calculations performed above and leave this more complicated
computation for further study.

\section{Gluon shadowing and the triple-Pomeron
diffraction}\label{gluons}

First of all, to avoid confusion it should be emphasized that we are not
talking about gluon shadowing in high-$p_T$ hadron production at $x_F=0$
in $d-Au$ collisions. This process exploits Bjorken $x>0.01$ which are
too large for gluon shadowing \cite{knst,kst2}. On the contrary, we
consider gluon shadowing in the soft inelastic $d-Au$ collisions which
are the main contributor to the total cross section. This process is
related to much smaller $x\sim 10^{-5}$.

Gluon shadowing is an important source of inelastic corrections at very
high energies. It is pretty clear if one employs Eq.~(\ref{54}). The part
of the diffraction which corresponds to the triple-Regge graph
$\Pom\Pom\Reg$, or the lowest order Fock component consisted only of
valence quarks, has a steep $M$-dependence, $d\sdd/dM^2 \propto 1/M^3$.
Therefore the integral over $M^2$ in (\ref{54}) well converges, the
minimal momentum transfer $q_L$ vanishes at high energies, and this part
of inelastic corrections saturates.

The triple-Pomeron ($\Pom\Pom\Pom$) part of diffraction which corresponds
to the Fock state containing at least one gluon, is divergent at large
masses, $d\sdd/dM^2 \propto 1/M^2$, since the gluon is a vector particle.  
The cut off is imposed by the nuclear formfactor in (\ref{54}), i.e. the
condition $q_L \lsim 1/R_A$. As a result of the divergence, this part of
the inelastic corrections rises as $\ln(s/s_0)$ and reaches a substantial
value at the energy of RHIC.

Eikonalization of the lowest Fock state $|3q\ra$ of the proton done in
(\ref{b.10})-(\ref{b.20}) corresponds to the Bethe-Heitler regime of
gluon radiation. Indeed, gluon bremsstrahlung is responsible for the
rising energy dependence of the phenomenological cross section
(\ref{170}), and in the eikonal form (\ref{b.10})-(\ref{b.20}) one
assumes that the whole spectrum of gluons is multiply radiated. However,
the Landau-Pomeranchuk-Migdal (LPM) effect \cite{lp,m} is known to
suppress radiation in multiple interactions. Since the main part of the
inelastic cross section at high energies is related to gluon
radiation, the LPM effect becomes a suppression of the cross section.
This is a quantum-mechanical interference phenomenon and it is a part of
the
suppression called Gribov's inelastic shadowing. The way how it is taken
into account in the QCD dipole picture is inclusion of higher Fock
states, $|3qG\ra$, etc. Each of these dipoles is of course colorless and
its elastic amplitude on a nucleon is subject to eikonalization.

As already mentioned, Eq.~(\ref{54}) should not be used at high
energies as it misses all higher order multiple off-diagonal transitions,
and incorrectly (ad hoc) calculates diagonal ones. On the other hand, the
eigenstate expressions Eqs.~(\ref{b.10})-(\ref{b.20}) is not safe to use
either. Indeed, the significant part of the integral over $M^2$ in
(\ref{54}), next to the upper cut off, corresponds to a finite $q_L$. In
other words, the fluctuation {\it valence quarks + gluons} is not frozen
by Lorentz time dilation during propagation through the nucleus.

\subsection{The Green function for glue-glue dipoles}\label{GF}

A proper treatment of a quark-gluon fluctuation "breathing" during
propagation through a nucleus is offered by the light-cone Green
function formalism.  In this approach the absorption cross section, as
well as the phase shifts, are functions of longitudinal coordinate. This
is also a parameter-free description, all the unknowns are fixed by the
comparison with other data. We employ this approach and calculate gluon
shadowing following Ref.~\cite{kst2}.

The key point which affects the further calculations is the
nonperturbative
light-cone wave function of the quark-gluon Fock state,
 \beq 
\Psi_{qG}(\vec r)= \frac{2}{\pi}\,
\sqrt{\frac{\alpha_s}{3}}\,
\frac{\vec e\cdot\vec r}{r^2}\, 
{\rm exp}\left(-{r^2\over2r_0^2}\right)\ . 
\label{3.1.24}
 \eeq
 Here we assume (as usual) that the gluon is carrying a negligible
fraction, $\alpha_G\ll1$, of the quark momentum. This wave function is
quite different from the perturbative one which is the same as in
light-cone description of the Drell-Yan process \cite{hir,kst2}. The
latter, employed for calculation of diffractive gluon radiation (the
triple-Pomeron term), results in overestimation of data for large mass
diffraction more than by order of magnitude. This problem has been known
since the 70s as the puzzle of smallness of the triple-Pomeron coupling.
The way out is to make a natural assumption that the parent light-front
quark and gluon experience a nonperturbative interaction which squeezes
that quark-gluon wave packet, and therefore reduces the dipole cross
section. The parameter $r_0$ in (\ref{3.1.24}) controls the strength of
the real part of the light-cone potential which is chosen in a Gaussian
form. Fit to diffractive data $pp\to pX$ leads to the value of the mean
transverse $q-G$ separation $\sqrt{\la r^2\ra}\equiv r_0=0.3\fm$. This
conclusions goes along with the results of nonperturbative models, like
the instanton vacuum model \cite{shuryak}, and lattice calculations
\cite{pisa}, which found a similar small size for gluonic fluctuations.
Such a semihard scale, $1/r_0$ also leads to quite a steep energy
behavior of the radiation cross section and well explains data for the
total and differential elastic cross sections of pp scattering
\cite{k3p}.

Apparently, smallness of $r_0$ leads to quite a weak shadowing for Fock
states containing gluons. As a consequence, we expect rather weak gluon
shadowing, what is not a surprise in view of the close connection between
diffraction and shadowing. As long as the gluon clouds around valence
quarks small, the Gribov's corrections are suppressed. Besides the
fluctuations containing gluons become heavy and the onset of gluon
saturation takes place at much smaller $x$ than usually expected.

The mean quark-gluon separation $r_0\approx 0.3\,fm$ is much
smaller than the quark separation in light hadrons. For this reason one
can neglect the interferences between the amplitudes of gluon radiation
by different valence quarks. Since the gluon contribution to the cross
section corresponds to the difference between the amplitudes of
$|qqqG\ra$ and $|qqq\ra$ components, the spectator quarks cancel out.
Then the radiation cross section is controlled by the quark-gluon wave
function and color octet ($GG$) dipole cross section.

Thus, the contribution to the total hadron-nucleus cross section which comes 
from gluon radiation has the form,
 \beq
\sigma^{hA}_G = \int\limits_x^1
\frac{d\,\alpha_G}{\alpha_G}\int d^2b\,
P(\alpha_G,\vec b)\ ,
\label{250}
 \eeq
 where $\alpha_G$ is the fraction of the quark momentum carried by the
gluon;
 \beqn
P(\alpha_G,\vec b) &=& 
T_A(b)\,\int d^2r\,
\Bigl|\Psi_{qG}(\vec r,\alpha_G)\Bigr|^2\,
\sigma_{GG}(r,s)
\nonumber\\ & -\, &
{1\over2}\,{\rm Re}\int\limits_{-\infty}^{\infty}
dz_1\,dz_2\,\Theta(z_2-z_1)\,
\rho_A(b,z_1)\,\rho_A(b,z_2)\int d^2r_1\,d^2r_2\,
\label{260}\\
&\times&
\Psi^*_{qG}(\vec r_2,\alpha_G)
\sigma_{GG}(r_2,s)\,
G_{GG}(\vec r_2,z_2;\vec r_1,z_1)\,
\sigma_{GG}(r_1,s)\,
\Psi_{qG}(\vec r_1,\alpha_G)\nonumber\ .
 \eeqn
 Here the energy and Bjorken $x$ are related as
$s=2m_NE_q=4/xr_0^2$. 

The second term in (\ref{260}) corresponds to the triple-Pomeron part of
the inelastic correction Eq.~(\ref{54}) written in impact parameter
representation.  The amplitude of diffractive gluon radiation $qN\to GqN$
is proportional to $\Psi_{qG}(\vec r,\alpha_G)\sigma_{GG}(r)$. A
glue-glue dipole emerges in this expression because this is not elastic
scattering, but a production process. Its amplitude comes from the
difference of the scattering amplitudes of different Fock components of
the quark \cite{hir}, $|q\ra$ and $|qG\ra$, which is a dipole cross
section of a color octet-octet dipole, $\bar qq-G$. Since the size of the
$\bar qq$ pair is irrelevant for gluon shadowing, we neglect it and
replace the $\bar qq$ by a gluon (see in \cite{kst1,kst2}). Therefore the
second term in (\ref{260}) can be interpreted as production of a $qG$
pair at the point $z_1$, then propagation of this pair with varying
transverse separation up to point $z_2$ where it converts back to the
quark.

Propagation of the dipole of varying size through the absorptive medium
between points $z_1$ and $z_2$ is described by the Green function
$G_{GG}(\vec r_2,z_2;\vec r_1,z_1)$. It satisfies the two dimensional
Schr\"odinger equation,
 \beq 
i\,\frac{d}{d\,z_2}\;
G_{GG}(\vec r_2,z_2;\vec r_1,z_1) =
\left[ - \frac{\Delta(\vec r_2)}
{2\,E_q\,\alpha_G(1-\alpha_G)} +
V(\vec r_2,z_2)\right]\,
G_{GG}(\vec r_2,z_2;\vec r_1,z_1)\ ,
\label{270}
 \eeq
 where imaginary part of the light-cone potential is related to
absorption in the medium,
 \beq
{\rm Im}\,V(\vec r,z) = - {1\over2}\,
\sigma_{GG}(\vec r)\,\rho_A(b,z)\ .
\label{280}
 \eeq
 For further calculations we assume that the quark energy is
$E_q=s/6m_N$, but the results are hardly sensitive to this approximation.

Perturbative calculations treating a quark-gluon fluctuation as free
particles overestimates the cross section of diffractive gluon radiation
(or the triple-Pomeron coupling) more than by an order of magnitude. The
only way to suppress this cross section is to reduce the mean transverse
size of the fluctuation. This is done in \cite{kst2} via introduction of
a real part of the light-cone potential in (\ref{270}),
 \beq
{\rm Re}\,V(\vec r,z) =
\frac{r^2}
{2\,E_q\,r_0^4\,\alpha_G(1-\alpha_G)}\ ,
\label{290}
 \eeq
  where parameter $r_0=$ was fitted to data for single diffraction $pp\to
pX$.
 
The gluonic dipole cross section $\sigma_{GG}(r,s)$ is assume to be
different from the $\bar qq$ one, Eq.~(\ref{170}), only by the Cazimir
factor $9/4$. To simplify calculations we rely on the small-$r$
approximation, $\sigma_{GG}(r,s)\approx C_{GG}(s)\,r^2$, where $C_{GG}(s)
= d\,\sigma_{GG}(r,s)/d\,r^2_{r=0}$. This approximation for the dipole
cross section is justified by the small value of $r_0^2\approx
0.1\,fm^2$.

In the case of a constant nuclear density,
$\rho_A(r)=\rho_A\,\Theta(R_A-r)$, the solution of Eq.~(\ref{270}) has a
form,
 \beqn 
&& G_{GG}(\vec r_2,z_2;\vec r_1,z_1) = 
\frac{A}{2\pi\,{\rm \sinh}
(\Omega\,\Delta z)} \nonumber\\
&\times& {\rm exp}\left\{-\frac{A}{2}\, 
\left[(r_1^2+r_2^2)\,{\rm coth}(\Omega\,\Delta z) - 
\frac{2\vec r_1\cdot\vec r_2}{{\rm sinh}
(\Omega\,\Delta z)} \right]\right\}\ , 
\label{300}
 \eeqn
 where
 \beqn A&=& {1\over r_0^2}\sqrt{1-
i\,\alpha_G(1-\alpha_G)\,E_q\,C_{GG}\,
\rho_A\,r_0^4}\nonumber\\ 
\Omega &=& \frac{i\,A}
{\alpha_G(1-\alpha_G)\,E_q}\nonumber\\ 
\Delta z&=&z_2-z_1\ .
\label{310}
 \eeqn

Integrations in (\ref{260}) can be performed analytically,
\beq
P(\alpha_G,\vec b) =
\frac{4\,\alpha_G}{3\,\pi}\,
\Re\ln(W)\ ,
\label{330}
\eeq
where
\beq
W = \cosh(\Omega\,L) +
\frac{A^2r_0^2+1}{2\,A}\,
\sinh(\Omega\,L)\ ,
\label{340}
\eeq
\beq
L=2\,\sqrt{R_A^2-b^2}\ .
\label{350}
\eeq

The first term in (\ref{260}) is a part of the nuclear cross section
calculated in Bethe-Heitler limit, i.e. without gluonic inelastic
shadowing. Therefore it is included in the nuclear cross sections
calculated so far. The new inelastic shadowing correction comes from the
second term in (\ref{260}). Its fraction of the total $pA$ cross
section is depicted in Fig.~\ref{glue}.
 \begin{figure}[tbh] 
\includegraphics{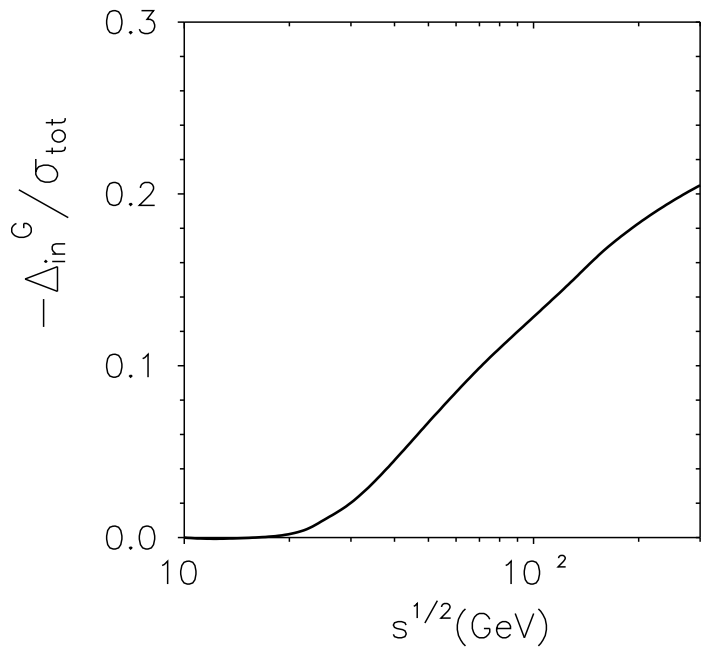} 
\begin{center} 
\vspace{9cm}
\parbox{13cm}
 {\caption[Delta] {\sl Ratio of the gluonic inelastic shadowing
correction (minimal bias) to the total nuclear cross section as function
of c.m. energy $\sqrt{s}$.}
 \label{glue}} 
\end{center}
 \end{figure}
 The onset of shadowing is delayed up to $\sqrt{s}\sim 20\GeV$. We
believe that this result is trustable since the Green function approach
treats phase shifts and attenuation in nuclear matter consistently.
Nevertheless, in order to get an idea about the scale of theoretical
uncertainty we also evaluated the magnitude of gluon shadowing using the
known values of the triple-Pomeron coupling and equation (\ref{54}). The
results are quite similar, in both cases the gluon shadowing correction
is pretty small \cite{kst2}, $\sim 20\%$ at the energy of RHIC. Such a
weak shadowing is a direct result of smallness of the parameter
$r_0=0.3\fm$ which we use. This seems to be the only way to suppress
diffractive gluon radiation corresponding to the triple-Pomeron
contribution, and to reach agreement with data on diffractive
dissociation $pp\to pX$. For this reason, all effects related to gluons,
including saturation, or color-glass condensate, are quite suppressed.

Naturally, the inelastic correction in (\ref{260}), (\ref{340}) varies
with impact parameter vanishing on the very periphery and reaching a
maximum at central collisions. At small $T_A(b)$ the inelastic correction
is proportional to $T_A^2(b)$ while the partial amplitude is proportional
to $T_A(b)$. Therefore, the ratio linearly rises with $T_A(b)$ (see in
\cite{knst1,krtj}) with a coefficient approximately equal to $0.2\fm^2$.
For very large $T_A^2(b)$ the correction may even exceed the rest of the
cross section, then apparently higher order corrections must be added to
stop this growth. Such a saturation is not important for real nuclei,
therefore we use the linear parametrization $R_G(b)=1-\Delta^G_{in}(b)=1
- 0.2\,T_A(b)$ for further calculations.

The valence quark part of the inelastic shadowing corrections makes the
nucleus more transparent, i.e. reduces the elastic scattering amplitude
as one can explicitly see comparing the corrected amplitude
Eq.~(\ref{60}) with the Glauber form Eq.~(\ref{a.50}). However, both
approach the black disc limit for large $T_A(b)\stn\gg1$. Important
question is whether this is still true after inclusion of gluonic
corrections.

Eq.~(\ref{260}) has the typical form of a nonlinear equation like
Glibov-Levin-Ryskin evolutions equation (GLR) \cite{glr}, or in the
dipole form Balitsky-Kovchegov equation (BK) \cite{bk}. The second
term on the right-hand side of (\ref{260}) corresponds to glue-glue
fusion in GRL or the multiple interaction in the nucleus in BK equations.
We calculated the correction in the lowest order using the uncorrected
dipole cross section $\sigma_{GG}(r)$, i.e. the undisturbed free gluon
density. Next iterations would be to implement the corrected gluon
density (at larger $x$, however), or $\sigma_{GG}(r)$, into the second
term in (\ref{260}). This procedure leads to the BK equation which
solution is still a challenge. However, due to smallness of the
correction, $20\%$, we do not expect large higher order
corrections and the saturated solution should not be very different from
our result which we employ in further applications.

On the other hand, if gluon shadowing emerging from the first order
iteration is very strong, like it was found in \cite{fs1,fs2,sarcevic},
it
should be substantially reduced by next iterations which effectively play
role of self-screening.  Namely, as long as the gluon density is reduced
at small $x$, one cannot use in Eq.~(\ref{54}) the cross section of
diffractive dissociation on a free nucleon target. It is suppressed by
the same gluon shadowing (at larger $x$ though). The stronger is the
gluon shadowing the more important is this self-screening effect. It was
missed in calculations \cite{fs1,fs2,sarcevic} which grossly
over-predicted the
strength of gluon shadowing.

Now we are in a position to correct our previous calculations for the
gluonic part of inelastic shadowing which we fix at $20\%$. 
We do it replacing $\sigma_{\bar qq}(r_T) \Rightarrow
R_G(b)\,\sigma_{\bar qq}(r_T)$, where $R_G(b) = 1-\Delta_{in}(b)$ is the
suppression factor related to gluon shadowing. This simple prescription
is based on the intuitive expectation that a dipole interacts with a less
number of gluons in the nucleus than the eikonal model assumes. Indeed,
for small separations $r_T$, the dipole cross section reads \cite{fs94},
$\sigma_{\bar qq}(r_T)=(\pi^2/3)\alpha_s\,r_T^2\,G(x,r_T)$, i.e. it is
indeed proportional to the gluon density which is reduced in nuclei.
More motivations for this procedure can be found in \cite{knst1,krtj}.

The results for nuclear cross sections corrected for gluon shadowing are
shown in Table~\ref{tab} and depicted in
Fig.~\ref{results-in}-\ref{b-neutron-in} by thick solid curves.

\subsection{More models for gluon shadowing}\label{models}

Although we predict quite a modest qluon shadowing effect and therefore a
rather small inelastic shadowing correction, many models predict much
stronger effects. One can call it theoretical uncertainty, if one treats
all models equally (though some of them are probably more equal than
others\cite{equal}). 

For instance, the popular event generator HIJING contains a
$Q^2$-independent gluon shadowing \cite{hijing} which is a factor $0.3$
at $x\sim 10^{-5}$. With such a dramatic gluon shadowing we get the
impact parameter dependence of the inelastic cross section depicted by
dotted curve in Fig.~\ref{results-in}. The corresponding correction
factor $K=0.65$ for the PHENIX data.

If to treat shadowing in terms of the dipole approach, it is clear that
shadowing is a monotonic function of $Q^2$, since the size of the dipole
can only rise towards the soft limit. This is confirmed by DGLAP evolution of
nuclear shadowing in the perturbative domain. Therefore, one can use
gluon shadowing predicted by different models at the starting scale $Q_0$
of the order of $1-2\GeV^2$ as a bottom bound for the shadowing
correction expected in the soft limit. We calculate Bjorken $x$ for the
RHIC energy $\sqrt{s}=200\GeV$ and $Q^2=1\GeV^2$.

A strong gluon shadowing was predicted in 
\cite{fs1,fs2}, $R_G=0.3-0.4$. HERA data for diffraction $\gamma^*p \to
Xp$ was used as an input in equation (\ref{54}) modified for $\gamma^*A$
collisions. The statistics of this data is much lower than in proton
diffraction $pp\to pX$ and not sufficient for reliable determination of
the triple-Pomeron coupling. Different solutions for this coupling fitted
to DIS diffractive data vary dramatically \cite{fit}. Besides, as is
mentioned above, the gluon self-screening missed in \cite{fs1,fs2} should
significantly reduce the effect of gluon shadowing.

Explicit calculations of gluon shadowing via gluon dipoles was performed
in \cite{sarcevic}. The gluon shadowing corresponding to the RHIC
energies was found at $R_G\approx 0.6$ what leads to correction factor
$K=0.78$ for the Phenix ratio $R_{dA}$. This calculation, however, also
does not include the gluon self screening, and is based on the assumption
that gluon and quark dipoles have identical distribution functions.

A strong gluon suppression was also found in a model with an early onset
of strong saturation \cite{levin} which characteristic scale is a steep
function of energy, $Q^2_s \propto (1/x)^{0.252}$. It was assumed in the
KLM approach \cite{klm} that for $Q^2 \leq Q^2_S$ gluon density
$xG_A(x,Q^2)$ is proportional to $Q^2\,R^2_A$ with a factor which was
taken from the McLerran-Venugopalan model \cite{mv} at $x = 10^{-1}$.
Such an oversimplified picture exhibits a strong gluon shadowing. If to
compare the $xG_A(x,Q^2)$ with the GRV parametrization \cite{grv} at
$x\sim 10^{-4}$ it turns out to be strongly suppressed by factor
$R_G=0.42$. In this case the correction factor in (\ref{6}) is $K=0.72$.

Such a diversity of model predictions suggests a conclusion that the
current data for deuteron-gold collisions \cite{phenix,star,phobos}
cannot resolve in a model independent way the dilemma, whether final
state interaction or initial conditions is the main source of hadron
suppression in heavy ion collisions. Indeed, if the latter were true, it
would unavoidably lead to a substantial reduction of $\sigma^{NA}_{in}$
and the ratio Eq.~(\ref{6}) (compared to the Glauber model).

\subsection{Number of participants}\label{npart}

Although the concept of number of participants originates from a naive
treatment of multiparticle production called wounded nucleon model, it is a widely 
used
characteristics of centrality of collisions.  We are not going to dispute
here its meaning, but just to see how it is affected by the inelastic
corrections\footnote{I am thankful to Larry McLerran suggested
to look at this parameter.} relying on its formal definition,
 \beq
\left.\frac{dN_p(s,b)}{d^2b}\right|_{Gl} =
T_A(\vec s-\vec b)\left\{1-
\exp\Bigl[-\sinn\,T_B(b)\Bigr]\right\} +
T_B(s)\left\{1-
\exp\Bigl[-\sinn\,T_A(\vec s-\vec b)\Bigr]\right\}\ ,
\label{360}
 \eeq
 where $\vec s$ is the parameter of collision of nuclei $A,B$.
We use subscript $Gl$ to emphasize that it corresponds to this model
which is inspired by the Glauber model (although they have nothing in 
common).

Apparently, inelastic shadowing corrections should reduce
$N_p$ since nuclear matter becomes more transparent. The corrected 
expression for $N_p$ reads,
 \beqn
\frac{dN_p(s,b)}{d^2b} &=&
\sinn\,T_A(\vec s-\vec b)\,T_B(b)
\nonumber\\ &\times& \left\{\frac{R_G(b)}{1+
R_G(b)\,\sinn\,T_B(b)} +
\frac{R_G(\vec s-\vec b)}
{1+R_G(\vec s-\vec b)\,\sinn\,T_A(\vec s-\vec b)}
\right\}\ .
\label{370}
 \eeqn
 Here the gluon shadowing factor $R_G$ is function of impart parameter
according to calculations in \cite{kst2,knst1,krtj} and the parametrization 
used above. 
We present the correction to the ''Glauber''
expression defines as,
 \beq
\delta_{shad}(b)= \frac{R_G(b)\,\sinn\,T_B(b)}{1+ 
R_G(b)\,\sinn\,T_B(b)}\left/\left\{
1-\exp\Bigl[-\sinn\,T_B(b)\Bigr]\right\}\right.\ ,
\label{380}
 \eeq
 in Fig.~\ref{npart-fig} depicted by solid curve.
 \begin{figure}[tbh] 
\includegraphics{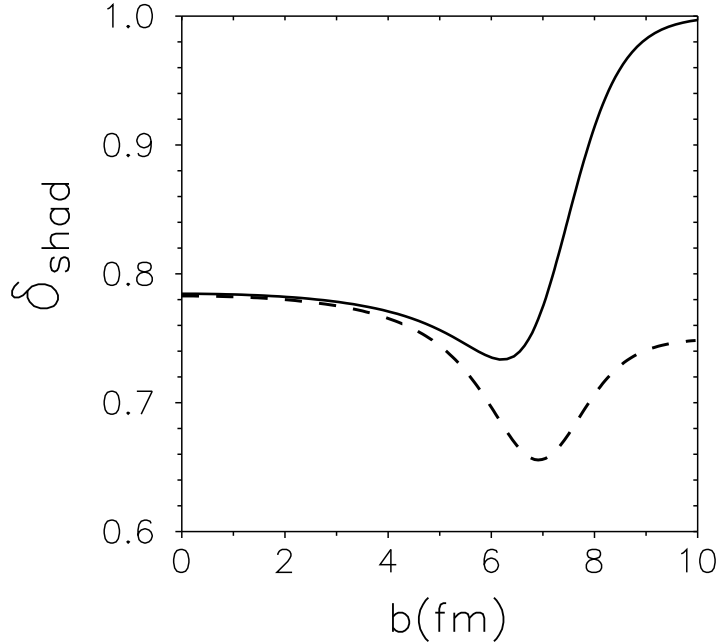} 
\begin{center} 
\vspace{9cm}
\parbox{13cm}
 {\caption[Delta]
 {\sl Solid line is the correction factor Eq.~(\ref{380}) for inelastic
shadowing to the number of participants in $p-Au$ collisions as function
of impact parameter. Dashed curve also includes a correction to $\sinn$
(see text).}
 \label{npart-fig}} 
\end{center}
 \end{figure}
 As one could expect, the correction factor peaks at the nuclear 
periphery and approaches one at large impact parameters.

If to compare with Glauber calculations employing the incorrect inelastic
cross section $\sinn=42\mb$, the correction is even larger, as is
demonstrated by dashed curve in Fig.~\ref{npart-fig}.

\section{Cronin effect: renormalizing the data}
\label{cronin}

Cronin effect for high-$p_T$ pions at $\sqrt{s}=200\GeV$ was predicted in
\cite{knst} to be a rather small enhancement, about $10\%$ at the maximum. 
The
smallness of the effect is due to the change of the mechanism of high-$p_T$
particle production which takes place at the RHIC energies. At lower
energies (SPS, CERN) different bound nucleons contribute to this hard
process incoherently. The nuclear enhancement is due to initial/final
state $p_T$-broadening of partons propagating through the nucleus. This
broadening should not be translated into a modification of the parton
distribution in the nucleus since $k_T$-factorization is broken
\cite{bbl}. At high energy an incoming light-cone fluctuation which
contains a high-$p_T$ parton is freed via coherent interaction with many
nucleons in the target. It turns out that such a coherent mechanism leads
to a weaker Cronin enhancement than the incoherent one. This is why
calculations \cite{wang,vitev} missing this effect of coherence, predict
a stronger Cronin effect.

The PHENIX data for neutral pions \cite{phenix} are depicted in
Fig.~\ref{cronin-data} by full points in comparison with the predicted
ratio
\cite{knst}. However, as it was stressed above, the normalization of the
data is based on Glauber model calculations which are subject to different
corrections, all of which have negative sign. As a result, the data
should be renormalized according to Table~\ref{tab} by multiplying the
experimental values by coefficient $K=0.83$. The corrected data are shown
by open circles. 
 \begin{figure}[tbh]
\includegraphics{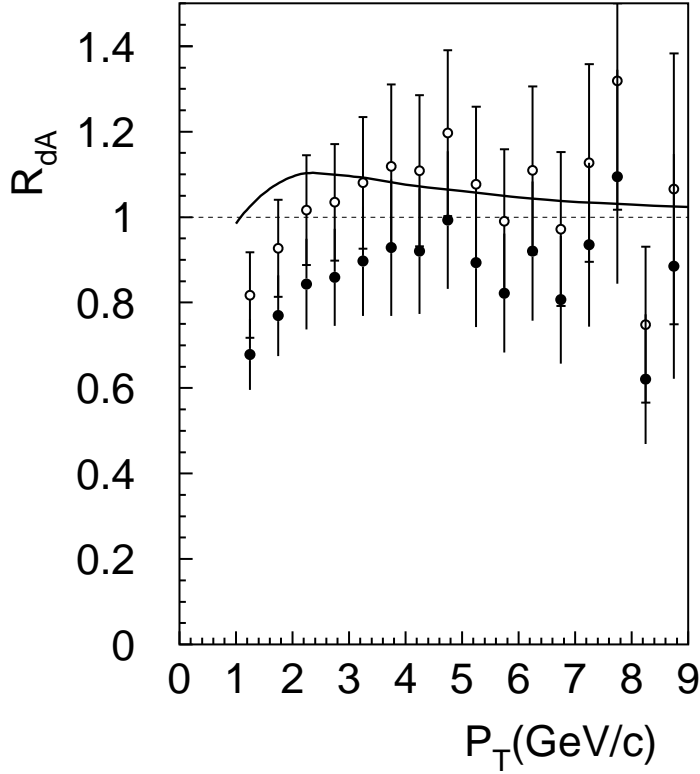}
 \begin{center}
 \vspace{11cm}
 \parbox{13cm} 
{\caption[Delta]
 {\sl The Cronin ratio $R_{Au/d}(p_T)$ for pions. Open circles show the
results of PHENIX with normalization base upon Glauber model calculations
of the inelastic $d-Au$ cross section using $\sinn=42\mb$
\cite{phenix}.  Full points show the same data corrected for a proper
value of inelastic $NN$ cross section and Gribov's inelastic shadowing.  
The error bars includes statistic and systematic uncertainties. The curve
is the prediction from \cite{knst}.}
 \label{cronin-data}} 
\end{center}
 \end{figure}

{\bf Cronin effect on a deuteron}

Theoretical predictions has been done so far for $pA$ collisions. In
order to compare models with $dA$ data one should make sure that
the Cronin enhancement on the deuteron itself is a small corrections.
We evaluate the ratio,
 \beq
R_{pd}(p_T)=
\frac{d\sigma^{pd}/d^2p_T}
{2\,d\sigma^{pp}/d^2p_T}\ ,
\label{390}
 \eeq
 in the limit of short coherence length which gives an upper  estimate
for the effect. Following \cite{knst} the $pd$ cross section at high
$p_T$ is given by  the following convolution,
 \beq
 \sigma_{pd}(p_T) = \sum\limits_{i,j,k,l}
 \widetilde F_{i/p} \otimes F_{j/d}
 \otimes \hat\sigma_{ij\to kl}\,
 \otimes D_{h/k}\ ,
 \label{400}
 \eeq 
 where $F_{i/p}$ and $F_{j/d}$ are the distributions of parton species
$i,j$ dependent on Bjorken $x_{1,2}$ and transverse momenta of partons in
the colliding proton and deuteron respectively.  The beam parton
distribution $\widetilde F^p_i$ is modified by the transverse momentum
broadening of the projectile parton due to interaction with another
nucleon in the deuteron. The broadening of the mean transverse momentum
squared reads \cite{dhk,jkt},
 \beq
\Delta\la k_T^2\ra = 2\,
\left.\frac{d\sigma_{\bar qq}(r_T)}
{dr_T^2}\right|_{r_T=0}\,\la T\ra\ ,
\label{410}
 \eeq
 where $\la T\ra$ is the mean nuclear (deuteron) thickness covered by the
projectile parton before or after the hard collision,
 \beq
\la T\ra = \frac{2}{\st}\int d^2s\, 
\Re\Gamma^{hN}(s)\,\Bigl|\Psi_d(s)\Bigr|^2\approx
\Bigl|\Psi_d(0)\Bigr|^2\ ,
\label{420} 
 \eeq
 where we neglected the elastic slope $B_{NN}$ compared to the nuclear
radius squared. For the parton distribution functions in a nucleon we use
the leading order GRV parametrization \cite{grv}.

 We calculated this ratio using the computer code for the Cronin effect
developed in \cite{knst}\footnote{I am thankful to Jan Nemchik who
performed this calculation for $pd$ collisions.}, and the deuteron wave
function $\Psi_d(\vec r_T)$ described in Appendix~\ref{appendC}.

The results for $R_{d/p}(p_T)$ are depicted in Fig.~\ref{deuteron}.
 \begin{figure}[tbh]
\includegraphics{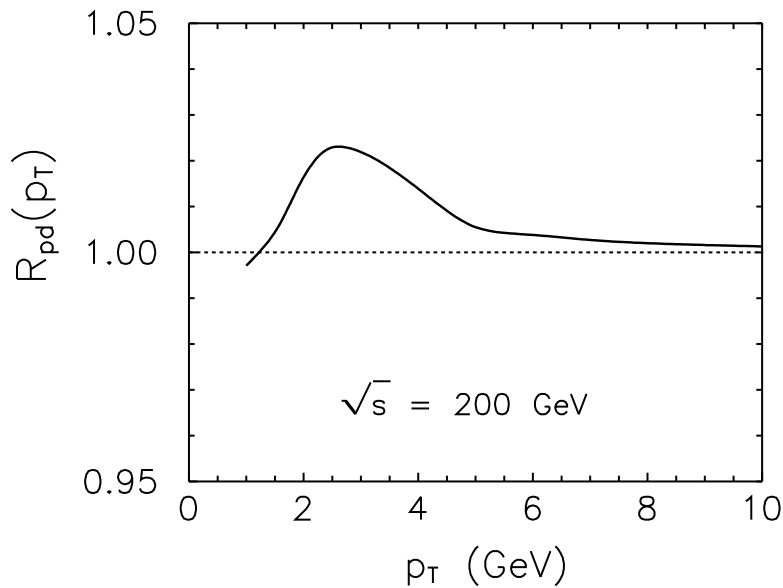}
 \begin{center}
 \vspace{8cm}
 \parbox{13cm} 
{\caption[Delta]
 {\sl Cronin ratio $R_{pd}(p_T)$ calculated at $\sqrt{s}=200\GeV$
using the formalism developed in \cite{knst}.}
 \label{deuteron}} 
\end{center}
 \end{figure}
 Indeed, the Cronin enhancement is only $2\%$, and can be neglected
comparing $d-Au$ data with predictions done for $p-Au$.

\section{Summary and conclusions}\label{summary}

The main observations and results of this paper are:
\begin{itemize}

\item
 The current normalization of inclusive high-$p_T$ cross section in
deuteron-gold collisions measured at RHIC is based on Glauber model
calculations of the inelastic $dAu$ cross section which is subject to
Gribov's inelastic shadowing corrections. Importance of these corrections
is not debatable, they have solid theoretical ground and are confirmed by
precise measurements \cite{murthy,gsponer} (see Fig.~\ref{murthy}). These
corrections, Eq.~(\ref{54}), have negative sign, i.e. make nuclear medium
more transparent, and they rise with energy.

\item
 First of all, the Glauber calculations must be improved.  The inelastic
$NN$ cross section used as an input should be corrected for diffraction.
For experiments insensitive to diffraction (PHENIX, PHOBOS) the cross
section should be reduced from $\sinn = 42\mb$ down to $\tilde\sinn\approx
30\mb$. On the contrary, if an experimental trigger detects diffraction
(STAR) this cross section should be increased up to $\stn=51\mb$. This 
modification results in a correction factor $K_{Gl}$ presented in 
Table~\ref{tab}.

\item There are two types of inelastic shadowing corrections. One
corresponds to diffractive excitation of the valence quark skeleton, or
nucleonic resonances, and is related to the $\Pom\Pom\Reg$ triple-Regge
graph. We calculated this correction, Eqs.~(\ref{60}), (\ref{145}),
(\ref{200}), using the light-cone dipole representation which effectively
sums up all orders of multiple interactions.

\item
 Another type of inelastic shadowing is related to diffractive gluon
bremsstrahlung, or to the soft limit of gluon shadowing in nuclei related
to the $\Pom\Pom\Pom$ triple-Pomeron diffraction. We performed
calculations in Sect.~\ref{gluons} using the solution for the Green
function, Eq.~(\ref{300}), describing propagation of a glue-glue dipole
through nuclear medium and found a rather weak gluon shadowing for gold,
about $20\%$. At the same time, other models predict much stronger gluon
shadowing ranging up to corrections of $70\%$ (Sect.~\ref{models}).

\item
 Altogether, we expect a reduction of inelastic $dAu$ cross section
compared to what was used for normalization of high-$p_T$ data at RHIC. We
conclude that the published data should be corrected by factor $K$ which
is about $0.8$ for PHENIX and about $0.9$ for STAR (see Table~\ref{tab}).
The renormalized data for pions do not possess any more the Cronin
enhancement. This correction factor might be even smaller, down to $0.65$
if to use a stronger gluon shadowing predicted by other models.

\item One should admit that current data for high-$p_T$ hadron production in
$dAu$ collisions at RHIC cannot exclude in a model independent way the
possibility of initial state suppression suggested in \cite{klm}, although
that would contradict the author's personal viewpoints.

\item Probably the only way to settle this uncertainty is a direct
measurement of either the cross section of high-$p_T$ pion production 
in $dAu$ collisions, or the inelastic $dAu$ cross sections at RHIC.

\item 
 A very sensitive test of models for inelastic shadowing offer tagged
events with a spectator nucleon.  In situation when direct measurement of
$dAu$ inelastic cross section is difficult, this might be a way to
restrict models and narrow the band of theoretical uncertainty. The
relative fraction of these events $20\%$ measured in \cite{star} create
apparent problems for models with strong gluon shadowing which predict a
much larger fraction. Even with our weak shadowing this fraction ranges
between $23\%$ and $26\%$. However, one should make it sure that the
detected neutrons are really spectators, what is not the case currently
(see discussion in Sect.~\ref{tagged}).

\item
 We found a beautiful quantum-mechanical effect: the nucleus acts like a
lens focusing spectators. In spite of the naive anticipation that nucleons
which escaped interaction retain their primordial Fermi momentum
distribution, there is a strong narrowing effect substantially reducing
the transverse momenta of the spectators. Besides, the distribution
acquires the typical diffractive maxima and minima (see
Figs.~\ref{q-tagg-minb} - \ref{q-tagg-central}).

\end{itemize}

\bigskip {\bf Acknowledgments:} This notes were written during visiting at
Columbia University and I am thankful to Miklos Gyulassy and Alberto Accardi
for hospitality and many inspiring discussions. I have also been much
benefited from discussions with Barbara Jacak, Peter Levai, Ziwei Lin, Sasha
Milov, Denes Molnar, Sergei Voloshin and other participants of the workshop,
as well as with Peter Braun-Munzinger, Claudio Ciofi, Alexei Denisov, J\"org
H\"ufner, Yuri Ivanov, Berndt M\"uller, Andreas Sch\"afer, Mike Tannenbaum,
and Xin-Nian Wang. My special thanks go to Yuri Ivanov and Irina Potashnikova
for their kind assistance with numerical calculations. This work is supported
by the grant from the Gesellschaft f\"ur Schwerionenforschung Darmstadt
(GSI), grant No.~GSI-OR-SCH, and by the grant INTAS-97-OPEN-31696.

\section{Appendix}

\def\appendix{\par
 \setcounter{section}{0}
 \setcounter{subsection}{0}
 \def\thesubsection{\Alph{subsection}}
 \def\theequation{\Alph{subsection}.\arabic{equation}}
 \setcounter{equation}{0}}

\appendix

\subsection{Glauber model glossary}\label{appendA}
\setcounter{equation}{0}

The $hA$ elastic amplitude at impact parameter $b$ has the eikonal 
form,
 \beq
 \Gamma^{hA}(\vec b;\{\vec s_j,z_j\}) =
1 - \prod_{k=1}^A\left[1-
 \Gamma^{hN}(\vec b-\vec s_k)\right]\ ,
 \label{a.10}
 \eeq
 where $\{\vec s_j,z_j\}$ denote the coordinates of the target nucleon
$N_j$. $i\Gamma^{hN}$ is the elastic scattering amplitude on a nucleon
normalized as,
 \beqn
\st &=& 2\int d^2b\,\Re\Gamma^{hN}(b);\nonumber\\
\sel&=& \int d^2b\, |\Gamma^{hN}(b)|^2\ .
\label{a.20}
 \eeqn
 
\subsubsection{Heavy nuclei}

In the approximation of single particle nuclear density one can calculate
a matrix element between the nuclear ground states.
 \beqn
\left\la0\Bigl|\Gamma^{hA}(\vec b;\{\vec s_j,z_j\})
\Bigr|0\right\ra =
1-\left[1-{1\over A}\int d^2s\,
\Gamma^{hN}(s)\int\limits_{-\infty}^\infty dz\,
\rho_A(\vec b-\vec s,z)\right]^A\ ,
\label{a.30}
 \eeqn
 where 
 \beq
\rho_A(\vec b_1,z_1) = \int\prod_{i=2}^A
d^3r_i\,
|\Psi_A(\{\vec r_j\})|^2\ ,
\label{a40}
 \eeq
 is the nuclear single particle density.

{\bf Total cross section}. 
The result Eq.~(\ref{a.30}) is related via unitarity to the total $hA$ 
cross section,
 \beqn
\sta&=&2\Re\int d^2b\,\left\{1 -
\left[1-{1\over A}\int d^2s\,
\Gamma^{hN}(s)\,T_A(\vec b-\vec s)\right]^A\right\}
\nonumber\\ &\approx&
2\int d^2b\, \left\{1-
\exp\left[-{1\over2}\,\st\,(1-i\rho_{pp})\,
T^h_A(b)\right]\right\}\ ,
\label{a.50}
 \eeqn
 where $\rho_{pp}$ is the ratio of the real to imaginary parts
of the forward $pp$ elastic amplitude;
 \beq
 T^h_A(b)= \frac{2}{\st}\int d^2s\, 
\Re\Gamma^{hN}(s)\,T_A(\vec b-\vec s)\ ;
\label{a.51} 
 \eeq
 and
 \beq
T_A(b) = \int_{-\infty}^\infty dz\,\rho_A(b,z)\ ,
\label{a.52}
 \eeq
 is the nuclear thickness function. 
 We use exponential form of $\Gamma^{hN}(s)$ throughout the paper,
 \beq
\Re \Gamma^{hN}(s) =
\frac{\st}{4\pi B_{hN}}\,
\exp\left(\frac{-s^2}{2B_{hN}}\right)\ ,
\label{a.53}
 \eeq
 where $B_{hN}$ is the slope of the differential $hN$ elastic cross
section. Note that the accuracy of the optical approximation in
(\ref{a.50}) is quite high for gold, $\sim 10^{-3}$, so we use it
throughout the paper. We also neglect the real part of the elastic amplitude
in what follows, since it gives a vanishing correction  $\sim 
\rho_{pp}^2/A^{2/3}$.

{\bf Elastic cross section}.
As far as the partial elastic amplitude is known, the elastic cross 
section reads,
 \beq
\sela=\int d^2b\, \left|1-
\exp\left[-{1\over2}\,\st\,T^h_A(b)\right]\right|^2\ .
\label{a.60}
 \eeq

{\bf Total inelastic cross section.} Apparently it is given by the
difference between the total and elastic cross sections,
\beq
\sigma^{hA}_{in} = \sta-\sela=
\int
d^2b\,\left\{1-\exp\left[-
\sigma_{tot}^{hN}\,T^h_A(b)\right]\right\}\ .
\label{a.65}
 \eeq
 This includes all inelastic channels when either the hadron or the
nucleus (or both) are broken up.

{\bf Quasielastic cross section}.
As a result of the collision the nucleus can be excited to a state 
$|F\ra$. Summing over final states of the nucleus and applying the 
condition of completeness, one gets the quasielastic cross section,
 \beqn
\sigma^{hA}_{qel} &=& \sum\limits_F
\int d^2b\ \left[\left\la0\left|
\Gamma^{hA}(b)\right|F\right\ra^\dagger
\left\la F\left|\Gamma^{hA}(b)\right|0\right\ra  -
 \left |\left\la 0\left|\Gamma^{hA}(b)
\right|0\right\ra\right|^2\right]\nonumber\\
&=& 
\int d^2b\ \left[\left\la0\left|
\Bigl|\Gamma^{hA}(b)\Bigr|^2
\right|0\right\ra
- \left |\left\la 0\left|\Gamma^{hA}(b)
\right|0\right\ra\right|^2\right]\ .
\label{a.70}
 \eeqn
 Here we extracted the cross section of elastic scattering when the
nucleus remains intact.

Then in the fist term of this expression we  make use of the relation,
\beq
\Re\int d^2s\,\frac{T^h_A(\vec b-\vec s)}
{A}\left\{1-2\Gamma^{hN}(s)+
\left[\Gamma^{hN}(s)\right]^2\right\}\approx
1-{1\over A}\,T^h_A(b)(\st-\sel)\ ,
\label{a.80}
 \eeq
 and arrive at,
 \beq
\sigma^{hA}_{qel} =
\int d^2b\,\left\{
\exp\left[-\sinhad\,T^h_A(b)\right]-
\exp\left[-\st\,T^h_A(b)\right]\right\}\ .
\label{a.90}
 \eeq

{\bf Inelastic nondiffractive cross section}. If one is interested in the
fraction of the total inelastic cross section (\ref{a.65}) which covers only
reactions with production of new particles, one should exclude the nucleus
break up to nucleons and nuclear fragments. That is the quasielastic cross
section, Eq.~(\ref{a.90}),
 \beq 
\sigma^{hA}_{prod} = \int d^2b\,
\left\{1-\exp\left[-\sinhad\,T^h_A(b)\right]\right\}\ . 
\label{a.100}
 \eeq
 This additional subtraction makes sense only for experiments which miss the
non-production break up of the nucleus. If, however, all inelastic events are
detected, including diffractive (production and non-production channels)
excitations of the nucleus (check with \cite{star}) one should rely on
Eq.~(\ref{a.65}) for the inelastic nuclear cross section.

{\bf Diffractive cross section}. One needs to know this cross section in
order to subtract it also from the inelastic cross section, since
diffractive events escape registration at $p(d)A$ collisions at SPS and
RHIC. The Glauber approximation is valid only for a single channel
problem. One can extend it to include diffraction properly introducing
phase shifts due to longitudinal momentum transfer. However, one needs to
know the cross section of interaction of the produced diffractive
excitation with nucleons. This goes beyond the reach of the Glauber
model, and instead of further ad hoc development of the model, we solve
this problem within the eigenstate method in Section~\ref{eigen}.

\subsubsection{Proton-deuteron collisions}

Apparently, Eq.~(\ref{a.50}) should not be applied to light nuclei, in
particular to a deuteron. Instead one should use,
 \beq
\std=2\stn + \Delta\std\ ,
\label{a.54}
 \eeq
 where
 \beq
\Delta\std = -2\int d^2b\int d^2r_T\,
\left|\Psi_d(r_T)\right|^2\,
\Gamma^{hN}(\vec b+\vec r_T/2)\,
\Gamma^{hN}(\vec b-\vec r_T/2)\ .
\label{a.55}
 \eeq
 One can switch via Fourier transform to momentum representation in each
of these three factors and perform integration over $\vec r_T$ and $\vec
b$. The result has a form of a one-dimensional integral \cite{glauber},
 \beq
\Delta\std =
- {2\over\pi}\int d^2q_T\,F_d(4q_T^2)\,
\frac{d\seln}{dq_T^2}\ ,
\label{a.58}
 \eeq
 where $F_d(q^2)$ is the charge formfactor of the deuteron. 
We neglected the correction $\sim 10^{-3}$ due to the nonzero real part
of the forward $NN$ amplitude. Note that $\vec s$ in (\ref{a.58}) is the
deuteron diameter, rather than the radius. This is why the formfactor
argument is $4q_T^2$.

We use
parametrization of the deuteron formfactor from \cite{anisovich},
 \beq
F_d(q_T^2) = 0.55\,e^{-\alpha q_T^2} + 0.45\,e^{-\beta q_T^2}\ ,
\label{30}
 \eeq
 and $\alpha=19.66\GeV^{-2}$, $\beta=4.67\GeV^{-2}$.

Using $\stn=51\mb$ at $\sqrt{s}=200\GeV$, and the elastic slope
$B_{NN}=14\GeV^{-2}$ we found the total $pd$ cross section, $\std=97\mb$
with Glauber correction $\Delta^{pd}_{Gl}=-5\mb$. Since at this point a
correct proton-deuteron cross section is needed, we have to go beyond the
Glauber approximation and add the inelastic correction considered
in Sect.~\ref{in-corr}. We show that it is equivalent to adding the
differential cross section of single diffraction, $pN\to XN$, to the
elastic one in (\ref{a.58}). This increases the value of the shadowing
correction by $1.75\mb$. Finally, we arrive at the cross sections,
 \beqn
\std &=& 95.15\mb\nonumber\\
\sind&=&\std - \frac{(\std)^2}{16\pi\,B_{pd}} = 84.9\mb\ .
\label{34}
 \eeqn
 Interesting, the inelastic cross section is not affected by the Glauber
correction, it is even slightly larger that the sum of two inelastic $NN$
cross sections.
 The slope from the differential elastic $pd$ cross section was measured and fitted in
\cite{akimov},
 \beq
\frac{d\sigma^{pd}_{el}}{dt} = \frac{(\sigma_{tot}^{pd})^2}
{16\pi}\ e^{B_{pd}t+C_{pd}t^2}\ ,
\label{24}
 \eeq
 where
 \beq 
B_{pd}=b_0 + b_1\ln s_{pd}\ ,
\label{26}
 \eeq
 with parameters $b_0=32.8\pm0.6\ (\GeV^2)$ and $b_1=1.01\pm0.09\
(\GeV^2)$. Parameter $C_{pd}=54.0\pm 0.9\ (\GeV^-2)$ was found to be
energy independent.  At the energy of RHIC $B_{pd} = 44.1\GeV^{-2}$, and we use this 
value in (\ref{34}).

The inelastic cross section Eq.~(\ref{34}) contains inelastic diffractive
channels like quasielastic break up of the deuteron, $pd\to ppn$, and
excitation of the nucleons $pd\to Xd$, $pd\to pY$, and $pd\to XY$.  For the
experiments insensitive to diffraction (PHENIX,PHOBOS) those channels must be
subtracted. 
 

\subsection{Deuteron wave function at rest and Lorentz
boosted}\label{appendC}
\setcounter{equation}{0}

To perform calculations for interaction of a high energy deuteron, one
should not use the three-dimensional deuteron wave function, but
needs to know the light-cone deuteron wave function expressed in Lorentz
invariant variables, the transverse $n-p$ separation $\vec r_T$
and the light-cone fraction $\alpha=p^+_n/p^+_d$ of the deuteron
momentum carried by a nucleon. One cannot get this wave function by a
simple Lorentz boost from the rest frame of the deuteron, where the
3-dimensional wave function is supposed to be known, to the infinite
momentum frame. Deuteron is not a classical system, under a Lorentz boost
it acquire new constituents which are quantum fluctuations. These
constituents build up higher Fock components. This makes the procedure of
Lorentz boost extremely complicated. There is, however, a practical
recipe suggested in \cite{terentev} and widely accepted. To the best
knowledge of the author, it works rather well for nonrelativistic
systems (nuclei \cite{fs-d}, heavy quarkonia \cite{hikt}, etc.)

The idea is straightforward, first, to express the deuteron wave function 
in momentum representation,
 \beq
\psi_d(\vec q) = \frac{1}{(2\pi)^3}
\int d^3r\,e^{i\vec q\cdot\vec r}\,
\psi_d(\vec r)\ ,
\label{a.110}
 \eeq
 via the light-cone variables in the rest frame of the deuteron. To do it
one should connect the three-dimensional momentum squared with the
effective mass of the $c\bar c$ pair, $q^2=M^2/4-m_N^2$, expressed in
terms of light-cone variables
 \beq
  M^2(\alpha,q_T) = \frac{q_T^2+m_N^2}{\alpha(1-\alpha)}\ .
 \label{a.120}
 \eeq
 In order to change integration variable $q_L$ to the light-cone $\alpha$
one uses their relation, $q_L=(\alpha-1/2)M(q_T,\alpha)$, and get a
Jacobian which can be attributed to the definition of the light-cone wave
function,
 \beq
  \psi(\vec q) \Rightarrow
  \sqrt{2}\,\frac{(q^2+m_N^2)^{3/4}}{(q_T^2+m_N^2)^{1/2}}
  \cdot \psi(\alpha,\vec q_T)
  \equiv \Psi(\alpha,\vec q_T) \ .
 \label{a.130}
 \eeq

Applying this procedure to the $S$ and $D$-wave radial wave functions one gets,
\beqn
\frac{u(\vec r)}{r} &\Rightarrow& U(\vec r_T,\alpha)\ ;
\nonumber\\
\frac{w(\vec r)}{r} &\Rightarrow& W(\vec r_T,\alpha)\ .
\label{a.140}
 \eeqn

This dependence on $\alpha$ is important for exclusive final states, for instance
deuteron dissociation to nucleons with definite longitudinal momenta. However, for most
of applications in this paper we need to know the $r_T$-retribution integrated over
$\alpha$,
 \beq
|\Psi_d(r_T)|^2 = \int\limits_0^1 d\alpha\,
\Bigl[U^2(r_T,\alpha)+W^2(r_T,\alpha)\Bigr]\ ,
\label{a.150}
 \eeq
 The result of this is identical 
to the simple integration over longitudinal variable in the rest frame of the nucleus,
 \beq
|\Psi_d(r_T)|^2 = \int\limits_{-\infty}^\infty dr_L\ 
\frac{u^2(r)+w^2(r)}{r^2}\ .
\label{a.160}
 \eeq

We use the contemporary deuteron wave functions which employ the
Nijmegen-93 potential \cite{nijm93}\footnote{I am grateful to Miklos
Gyulassy for suggesting this and providing relevant data.}.

\end{document}